\documentclass[english]{article}
\usepackage[T1]{fontenc}
\usepackage[utf8]{inputenc}
\usepackage{geometry}
\usepackage{color}
\usepackage{babel}
\usepackage{float}
\usepackage{amsmath}
\usepackage{amsthm}
\usepackage{amssymb}
\usepackage{graphicx}
\usepackage{setspace}
\PassOptionsToPackage{normalem}{ulem}
\usepackage{ulem}
\usepackage[unicode=true,pdfusetitle,
 bookmarks=true,bookmarksnumbered=false,bookmarksopen=false,
 breaklinks=false,pdfborder={0 0 1},backref=false,colorlinks=false]
 {hyperref}
\usepackage{breakurl}

\makeatletter

\providecommand{\tabularnewline}{\\}

\theoremstyle{definition}
 \newtheorem{example}{\protect\examplename}
\theoremstyle{plain}
\newtheorem{prop}{\protect\propositionname}
\theoremstyle{plain}
\newtheorem{thm}{\protect\theoremname}
\theoremstyle{plain}
\newtheorem{lem}{\protect\lemmaname}
\theoremstyle{plain}
\newtheorem{cor}{\protect\corollaryname}
\theoremstyle{remark}
\newtheorem{rem}{\protect\remarkname}

\makeatother

\providecommand{\corollaryname}{Corollary}
\providecommand{\examplename}{Example}
\providecommand{\lemmaname}{Lemma}
\providecommand{\propositionname}{Proposition}
\providecommand{\remarkname}{Remark}
\providecommand{\theoremname}{Theorem}

\begin{document}
\begin{doublespace}
\begin{center}
\textbf{\Large{}Securities Lending Strategies: Exclusive Valuations
and Auction Bids}{\Large\par}
\par\end{center}

\begin{center}
\textbf{Ravi Kashyap}
\par\end{center}

\begin{center}
\textbf{City University of Hong Kong / SolBridge International School
of Business}
\par\end{center}

\begin{center}
\par\end{center}

\begin{center}
and Prediction Methods / Simulation Methods 
\par\end{center}

\begin{center}
AMS Subject Codes: 91G20 Derivative securities; 91G60 Numerical methods;
60G25 Prediction theory
\par\end{center}
\end{doublespace}

\begin{center}
\begin{center}
\today
\par\end{center}\tableofcontents{}
\par\end{center}

\section{\quad Abstract}

We derive valuations of a portfolio of financial instruments from
a securities lending perspective, under different assumptions, and
show a weighting scheme that converges to the true valuation. We illustrate
conditions under which our alternative weighting scheme converges
faster to the true valuation when compared to the minimum variance
weighting. This weighting scheme is applicable in any situation where
multiple forecasts are made and we need a methodology to combine them.
Our valuations can be useful either to derive a bidding strategy for
an exclusive auction or to design an appropriate auction mechanism,
depending on which side of the fence a participant sits (whether the
interest is to procure the rights to use a portfolio for making stock
loans such as for a lending desk, or, to obtain additional revenue
from a portfolio such as from the point of view of a long only asset
management firm). Lastly, we run simulations to establish numerical
examples for the set of valuations and for various bidding strategies
corresponding to different auction settings.

\section{\label{sec:Introduction}Introduction}

Existing studies on securities lending fail to consider the full extent
to which lending desks bridge the demand and supply gap by setting
loan rates and managing inventory by finding securities externally
or using the positions of other trading desks within the same firm
(section \ref{sec:Related-Literature} summarizes much of the existing
literature in this space, from which we see that the actions of the
lending desks are mostly ignored; appendix \ref{sec:Securities-Lending-Background}
provides an introductory background to securities lending). A more
complete study on the effects of short selling must look to incorporate
the actions of the main players and how they look to alter their cost
structure or the demand/supply mechanisms by pulling the above levers
(setting loan rates and managing inventory) they have at their disposal.
In this paper and related works (Kashyap 2017), we derive various
theoretical results that consider the modus operandi of the players
in the lending space and supplement the results with practical considerations
that can be operationally useful on a daily basis.

The securities lending business is a cash cow for brokerage firms.
(D’Avolio 2002; Jones \& Lamont 2002; and Duffie, Garleanu, \& Pedersen
2002) have details on the mechanics of the equity lending market;
(End notes \ref{enu:One} and \ref{enu:Two}) have further details
on the historical evolution of securities lending. Lenders are assured
of a positive spread on every loan transaction they make. Historically,
the loan rates were determined mostly as a result of a bargaining
process between parties taking the loan and traders on the securities
lending desks. Recent trends due to increased competitive pressures
among different players (lending desks and other intermediaries),
the introduction of various third party agents that provide information
and advice to beneficial owners (the actual asset owners who supply
inventory to the lending desks), and the treatment of securities lending
as an investment management and trading discipline, have compressed
the spreads (difference between the rate at which lending desks acquire
inventory and the rate at which they make loans) and forced lending
desks to look for ways to improve their profit margins.

To aid this effort at profitability it is possible to: 1) develop
different models to manage spreads on daily securities loans and aid
the price discovery process; 2) improve the efficiency of the locate
mechanism and optimize the allocation of inventory; 3) develop strategies
for placing bids on exclusive auctions; 4) price long term loans as
a contract with optionality embedded in them; 4) look at ways to benchmark
which securities can be considered to be more in demand, or highly
shorted, and use this approach to estimate which securities are potentially
going to become “hot” or “special”, that is securities on which the
loan rates can go up drastically and supply can get constrained. (Kashyap
2016) looks at some of these recent innovations being used by lending
desks and also considers how these methodologies can be useful for
both buy side and sell side institutions (that is, for all the participants
involved).

\textbf{\textit{In this paper, we derive valuations of a portfolio
of financial instruments from a securities lending perspective. This
valuation would reflect the value of the portfolio of securities if
a certain percentage of the holdings had to be borrowed to cover corresponding
short positions. The valuation exercise then becomes an effort at
finding an annual rate to be paid to the actual owner of the portfolio
of securities that are being offered as an exclusive. This value represents
the fees that the participant getting the use of the exclusive (usually
securities lending desks) hopes to earn by lending out the securities
to their final end borrowers.}}

In the rest of this section (\ref{subsec:Securities-Lending-Exclusive}),
we discuss the structure of exclusive auctions and the mechanics of
how they are carried out, including the use of visual aids to better
illustrate the organization and interaction between the main players.
Section \ref{sec:Motivation-for-Exclusive} provides a deeper discussion
of the motivation for exclusive auctions and the benefits it provides
for the main participants. Section \ref{sec:Motivation-for-Exclusive}
also reviews many related papers in the securities lending literature
where the bulk of the focus has been to study the effects of short
selling on stock prices. This illustrates the complementary nature
of our paper since we focus on understanding one of the means through
which securities lending desks try to obtain inventory (that is securities
lending exclusives) and provide tools to come up with a valuation
for the same. Section \ref{sec:Exclusive-Valuation} develops the
theoretical and modeling aspects of an exclusive and provides the
main valuation results. Section \ref{sec:Auction-Strategy} summarizes
the main auction theory results (which are collected from other papers)
that are necessary to make an exclusive auction bid. This is done
so that readers will find this paper as a complete reference on securities
lending exclusives. As we clarify in the next section (\ref{subsec:Securities-Lending-Exclusive}),
before making an auction bid participants come up with a valuation
and then shade the valuation to suit the type of auction they are
participating in. Section \ref{sec:Numerical-Results} has numerical
results based on simulations and great care has been taken to ensure
that the results are close to what would be observed in an actual
exclusive auction. Sections \ref{sec:Improvements-to-the} and \ref{sec:Conclusion}
have suggestions for improvements, extensions that are possible to
the benchmark models derived here, assumptions that could be relaxed
to incorporate more complex models and the conclusion.

\subsection{\label{subsec:Securities-Lending-Exclusive}Securities Lending Exclusive
Structure}

\begin{figure}[H]
\includegraphics[width=17cm]{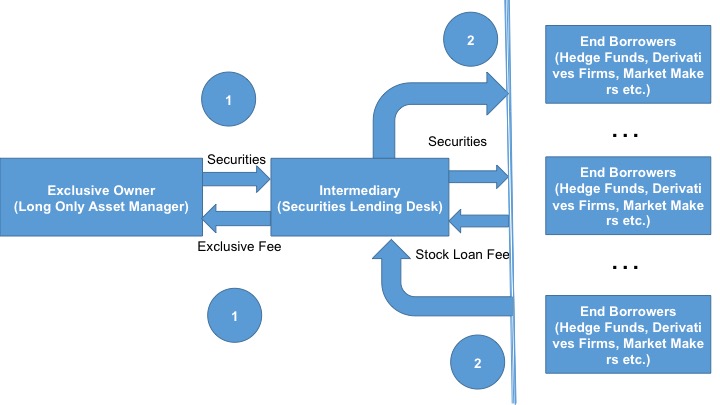}

\caption{\label{fig:Securities-Lending-Exclusive-Arrangement}Securities Lending
Exclusive Arrangement}
\end{figure}

(Figure \ref{fig:Securities-Lending-Exclusive-Arrangement}) is a
typical exclusive arrangement showing how an intermediary (securities
lending desk) sits between an exclusive owner (long only asset manager)
and final end borrowers (hedge funds, derivative traders, market makers,
etc.) who have short positions. The first portion of the figure (near
the circle marked one) shows the exclusive contract arranged between
the intermediary and the long only asset manager. This contract allows
the intermediary exclusive use of the holdings of the asset manager
for making stock loans. In return for being able to use the holdings
of the asset manager, the intermediary pays an annual fee. The contract
is usually made for a year (or multiple years in some instances) and
the fees are fixed when the contract is initiated. The second half
of the figure (near the circle marked two) shows the securities from
the exclusive portfolio being used by the intermediary to make loans
to the final end borrowers. The final end borrowers pay the stock
loan fee or the borrow costs; though the intermediary can change this
stock loan fee on a daily basis. The stock loans to end borrowers
are usually shorter in duration compared to the exclusive contract. 

\begin{figure}[H]
\includegraphics[width=17cm]{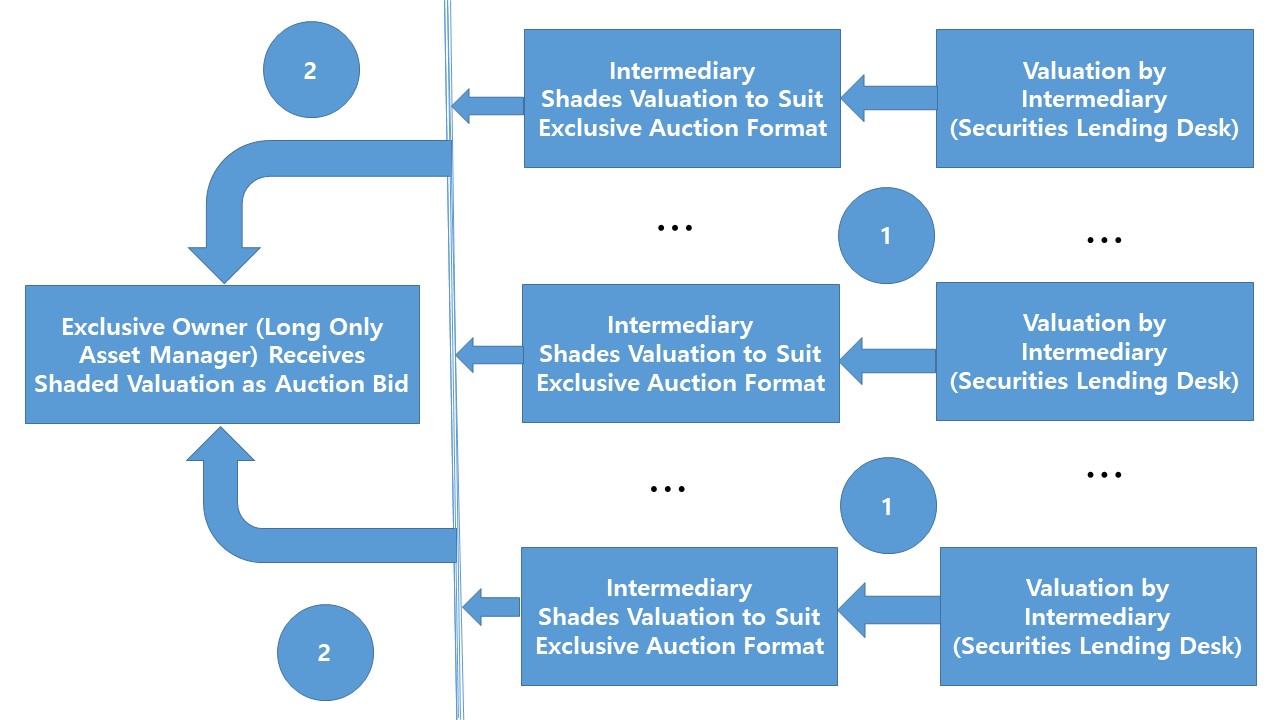}

\caption{\label{fig:Securities-Lending-Exclusive-Procurement}Securities Lending
Exclusive Procurement}

\end{figure}

The first step (Figure \ref{fig:Securities-Lending-Exclusive-Procurement};
near the circle marked one) in the determination of the fixed fee
paid by the intermediary to the exclusive portfolio owner is the valuation
of the portfolio. Section \ref{sec:Motivation-for-Exclusive} has
details on the motivation for procuring such an exclusive contract
and section \ref{sec:Exclusive-Valuation} considers the valuation
methodologies to arrive at a basis point estimate including various
assumptions that would be realistic from a securities lending point
of view. The long only asset manager (or an agent of the asset manager)
conducts an auction where many interested intermediaries are invited
to place bids for the holdings of the asset manager. Hence the valuation
obtained in the first step is shaded to suit the many possible auction
formats (Figure \ref{fig:Securities-Lending-Exclusive-Procurement};
near the circle marked two), which becomes the bidding strategy discussed
in section \ref{sec:Auction-Strategy}. The winner of the auction
is awarded the exclusive contract and the fixed fee is based on the
winning bid. Sometimes, the holdings are distributed among a few of
the top bidders and many rules are used to decide what fees apply
and how much of the holdings each bidder gets. It helpful to think
of the valuation of the portfolio as the initial step towards obtaining
an auction bidding strategy which is the ultimate goal of the auction
participants.

\begin{example}
\textbf{\textit{\label{Example-Portfolio-Value-Exclusive-Rate}}}For
example, if a portfolio with notional value 1 Billion USD was being
offered as an exclusive contract via an auction. If a certain sell-side
participant wishes to use this portfolio to lend to their end clients,
he would estimate an annual rate based on, among other things, the
securities lending loan rates of all the securities in that portfolio.
Let us say the estimate turns out to be 25 basis points and the lending
desk makes this their bid (without changing the valuation) and wins
the auction, the lending desk would then pay 2.5 million USD (1 Billion
times 25 basis points) to the exclusive portfolio owner. This example
has many simplifying assumptions which are relaxed in the corresponding
sections below.

An instance of this situation is when a large sovereign wealth fund
sells exclusive securities lending rights to a broker-dealer. A broker-dealer
may want to negotiate this type of contract to differentiate its service
to client hedge-funds. The assets for which exclusive access is sought
are typically difficult to trade, such as securities form a developing
countries, or desirable as a bundle, such as the securities underlying
an index. Hence, it becomes important to study the problem of using
an auction to value exclusive access rights to a block of securities
held by a buy-and-hold investor.
\end{example}
Such a valuation of any portfolio of financial instruments can be
useful either to derive a bidding strategy for an exclusive auction
or to design an appropriate exclusive auction mechanism, depending
on which side of the fence a participant sits (whether the interest
is to procure the rights to use a portfolio for making stock loans
such as for a lending desk, or, to obtain additional revenue from
a portfolio such as from the point of view of a long only asset management
firm). The valuation of the portfolio being auctioned is subject to
the information set available to the bidder or the auction designer.
This information set would include among other things: the demand
for the securities (say for example, shares being borrowed to short
an equity instrument); any additional demand from the loan locates
received (section \ref{sec:Motivation-for-Exclusive} provides an
explanation of the locate mechanism); the loan rates applicable to
those securities; the duration of the loans; the frequency of loan
turnover; the prices of the securities (stock price if the financial
instrument is equity); and the internal inventory pool available to
the bidder.

\textbf{\textit{Given this scenario, the exclusive valuation can be
viewed as an exercise to arrive at the properties of an entire portfolio
(the macroscopic system) after factoring the many random characteristics
of the individual securities in that portfolio (microscopic constituents).}}
Some of the variables can be modeled as Geometric Brownian Motions
(GBMs) with uncertainty introduced via suitable log-normal distributions
and certain others can be represented using folded asymmetric normal
distributions or by taking the absolute value of a normal distribution.
\textbf{\textit{This modeling approach allows the use of simulations
to value an exclusive contract making it comparable to the pricing
of a derivative contract. An exclusive (to be more precise, the revenue
that can be derived from an exclusive) can then be viewed as a contingent
claim on a portfolio, which enables the application of suitable option
pricing techniques.}}

Different assumptions regarding the different variables would lead
to different valuations. We derive heuristics to arrive at a set of
valuations, with a pecking order that can help decide the aggressiveness
regarding which of the valuations to chose from.\textbf{\textit{ A
key result (Theorem \ref{Th-1-When-each-of}) is a way to combine
different valuations such that the aggregated valuation asymptotically
arrives at the true value. This result is applicable to any problem
where multiple forecasts are derived and we need a methodology to
combine them. We show conditions under which our alternative weighting
scheme converges faster to the true valuation when compared to the
minimum variance weighting. }}

Once a valuation has been obtained, it is important to come up with
the best strategy from an auction perspective. \textbf{\textit{We
use existing and well known auction theory results along with extensions
from a related paper (Kashyap 2018).}} We start with the benchmark
scenario where the buyers placing bids are assumed to have perfect
and complete information, that is private only to them, regarding
their valuation of the portfolio that is being auctioned. We consider
the uniform distribution as the simplest scenario and extend that
to a more realistic setting that considers the valuations to be log
normally distributed. We further extend this by introducing uncertainty
into the estimation of bidder valuations and their bidding strategy.
The possibility of the number of bidders being unknown, the valuations
from various bidders being correlated or the interdependent valuation
framework and, a reserve price set by the auction seller are more
complex extensions. Based on existing results, it is easily seen that
the strategies of the bidders constitute a Nash equilibrium under
suitable conditions.

Lastly, we run simulations to establish numerical examples for the
set of valuations and for various bidding strategies corresponding
to the different auction settings. The next generation of models and
empirical work on securities lending activity would benefit by factoring
in the methodologies considered here. Understanding the inner workings
of securities lending players, including the provision of better tools
and models to aid their efforts, could counter the recent concerns
about risks in the securities lending space (End-note \ref{enu:Four}).
In addition, the models developed here could be potentially useful
for inventory estimation and for wholesale procurement of financial
instruments and also non-financial commodities.

From the discussion thus far and as the paper deveops, it will become
clear that exclusive valuation is an extremely challenging problem
in terms of the many sources of uncertainty that it holds. Having
better ways to manage this complexity, as illustrated in this paper,
will be well rewarded since this is one of the least explored yet
profit laden areas of modern investment management. For completeness,
we provide a brief motivation for exclusive auctions before delving
further into the mechanism of estimating an auction bid for exclusives.

\section{\label{sec:Motivation-for-Exclusive}Motivation for Exclusive Auctions}

We can trace the origins of Exclusive Auctions to the early 2000s.
(Duffie, Garleanu \& Pedersen 2002) briefly mention an exclusive lending
deal between Credit Suisse First Boston (CSFB) and California Public
Employees Retirement System (CalPERS) in 2000. We could found any
other reference on this topic in a serious academic paper. As with
the rest of the securities lending industry, this practice is more
prevalent for equity portfolios. As opposed to traditional arrangements
between intermediary brokers and beneficial owners, where the loan
rates on each security are negotiated periodically, an exclusive auction,
as the name suggests, provides sole usage of a portfolio of securities,
or to a portion of the portfolio, to the winner in an auction process
for a certain time period. This arrangement is beneficial to both
parties since the intermediary broker gets single ownership to the
portfolio. Intermediaries can use the portfolio as part of their overall
supply and even if the loan rates for a group of securities in the
portfolio go up, the costs of sourcing these special stocks remains
the same. Intermediaries look at exclusives as a source of locking
up inventory for a certain time horizon. Beneficial owners get a guaranteed
source of revenue and will not have the administration hassle of having
to constantly create new loans. They will not have to deal with multiple
intermediaries and can place their portfolio with an auction agent.
Both parties do not need to negotiate or renegotiate loan rates on
individual securities for the duration of the exclusive contract.

The holdings in the portfolio on certain key dates are provided to
the intermediary brokers or the agent administering the auction to
enable brokers to estimate the value of the portfolio from a lending
perspective. The intermediary brokers shade this valuation of the
portfolio to suit the auction mechanism and make bids accordingly.
The bid is usually expressed as a certain number of basis points of
the portfolio value at the time of auction, applicable annually or
over the duration of the exclusive agreement. In addition to the exclusive
bid, beneficial owners also sometimes charge transaction fees each
time securities are taken out from the portfolio or added back.

Beneficial owners continue to manage their portfolio positions as
per their investment mandates or according to their re-balancing guidelines
or risk tolerances. This risk of turnover in the holdings is something
that intermediaries need to factor in their exclusive bids. The agreements
can stipulate certain criteria on the turnover of the holdings, which
would require the exclusive fee to be reassessed. The huge size of
the portfolios that are generally auctioned and the relatively small
price of the exclusive fees, in comparison with the loan rates on
individual securities, mean that winning an auction bid is an extremely
profitable venture for intermediaries. In addition, by gaining access
to an exclusive portfolio, intermediaries prevent competitors from
having access to this source of inventory, almost acting like monopolists
in supplying loans for certain hard to borrow instruments. This restricted
supply enables loans to be priced higher. This phenomenon is partly
offset when a portfolio is auctioned to more than one bidder, but
still provides pricing power to the winners of the auction. This practice
of spreading very large portfolios across three or four of the top
bidders is becoming more common.

Sometimes, the lending desk could have access to shares available
to others parts of the firm when it acts as a primer broker, operates
derivative trading, proprietary trading or services private client
accounts. This additional inventory is readily available to the firm
as a side effect of having other business units or trading desks.
The lending desks at various firms are expected to fully utilize this
internal inventory before looking outside for additional supply. Complete
utilization of this internal inventory would reduce the funding costs
for the other business units and also make the loan rates charged
by the firm cheaper than the loan rates of other lending desks, when
it has significant internal inventory. The variation in the valuation
of the exclusive across different firms would then primarily depend
on the extent of the overlap of this internal supply with the holdings
in the exclusive. The other source of variation would be the loan
rates the lending desk applies to the loans it makes. Historically,
the loans rates across different lending desks of different intermediaries
have varied considerably due to the opaque nature of the transactions
and the variable demand seen by individual desks. With centralized
platforms, which consolidate and disclose rates across firms, coming
into vogue loan rates have converged to a considerable extent.

Another piece of the puzzle is the locate requests received by the
lending desk on a daily basis. These locate requests are sent by end
borrowers, in advance of actually borrowing shares to short, to get
an indication of the quantity of shares they can borrow. This is done
to ensure that their shorting needs for the trading day can be met.
The intermediary can fill either a portion or the entire locate request
depending on its inventory situation and also depending on how many
firms are sending it locates for that particular security for that
trading day. But once a locate request is filled by a lending desk,
they are expected to have that number of shares ready for the borrowing
firm. A borrowing firm, on the other hand, can borrow as much of the
filled locate amount as it chooses to. This mismatch between locate
approvals and actual borrows then leads to another aspect of the lending
business that can be optimized by implementing different variations
of the Knapsack Algorithm (Martello \& Toth 1987) and we consider
this in another paper (Kashyap 2016). The conversion factor from locates
to borrows can be estimated as part of the locate approval optimization.
For the present purpose of estimating an exclusive value, we take
this conversion factor as exogenously given. Lending desks have been
considering charging a nominal fee based on the locate amount they
agree to fill to discourage borrowers from sending in spurious locate
requests, though this practice is yet to be formally institutionalized
across the lending industry.

So in effect, the lending desk has a certain amount of borrows on
the book at any time, which is matched by a combination of internal
inventory and supply from beneficial owners. Excess demand arrives
in the form of locate requests. Existing loan borrowers can increase
their loan holdings via telephone or email, so the loan book can change
without the means of locate requests. Managing loan turnover, returning
or acquiring supply, locate fulfillment and negotiating the loan rates
then constitute the primary loan management duties of the desk. 

\subsection{\label{subsec:Exclusive-Auctions-Wallet}Exclusive Auctions Wallet
Size}

\begin{figure}[H]
\includegraphics{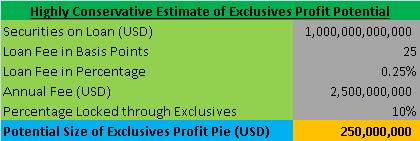}\caption{Exclusive Auctions Profit Potential Estimate\label{fig:Exclusive-Profit-Potential}}
\end{figure}

\begin{figure}[H]
\includegraphics[width=10cm]{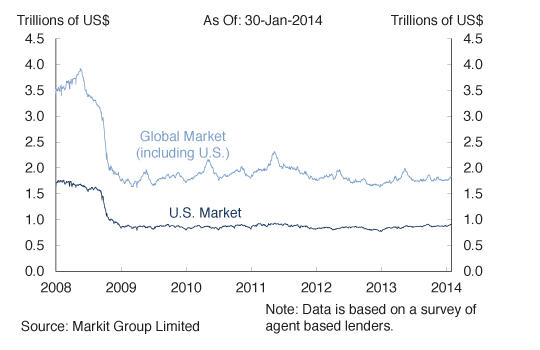}\textbf{\uline{USA}}\includegraphics[width=7cm]{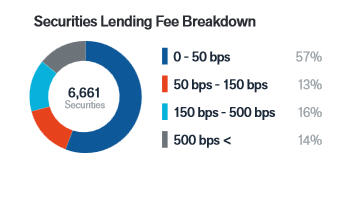}

\textbf{\uline{Asia}}\includegraphics[width=7cm]{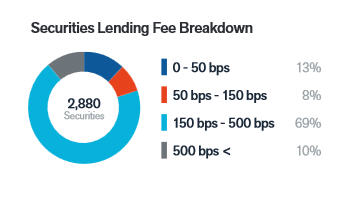}\textbf{\uline{Western
Europe}}\includegraphics[width=7cm]{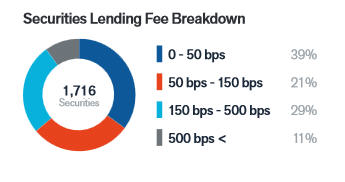}

\caption{Securities Lending Market Size and Loan Fees\label{fig:Securities-Lending-Market}}
\end{figure}

A rough estimate of the potential profits that could be accrued by
indulging in exclusives is shown in Figure \ref{fig:Exclusive-Profit-Potential}.
The point to keep in mind is that this is a highly conservative and
approximate estimate since we have used 1 Trillion USD as the notional
amount of securities on loan and around 25 basis points as the loan
fees in our estimate (End-note \ref{enu:Four} mentions that the volume
weighted average loan rate in the US is around 38 basis points; many
other markets have much higher weighted average loan rates). The global
securities on loan is around 2 trillion USD (Figure \ref{fig:Securities-Lending-Market})
and there are securities with loan rates of almost 25\%. (Baklanova,
Copeland \& McCaughrin 2015; End-notes \ref{enu:Three}, \ref{enu:Four})
have more details on the size of the securities lending market. Even
this simple back of the envelope calculation demonstrates that better
techniques could go a long way in boosting profits in the exclusive
auction process towards which, to the best of our knowledge, no prior
work has been done that applies the use of quantitative methodologies.

\subsection{\label{subsec:Buy-Side-and}Buy Side and Sell Side Perspective}

The sell side here would be the collection of intermediary firms that
source supply and lend it on to final end borrowers. The buy side
here would have two segments of firms. One, the end borrowers who
either have a proprietary trading strategy or hedging that requires
shorting certain securities. Two, the beneficial owners who are long
and provide supply to the intermediaries also fall under the buy side
category. Depending on which side a firm falls under, they will find
the below derivations useful, since it will affect the rates they
charge or the rates they pay. These methodologies will also help auction
designers, who operate on behalf of beneficial owners, formulate an
appropriate mechanism that results in the best outcomes for their
clients. This can provide transparency to the beneficial owners in
terms of how the actual valuation of the portfolio might differ from
the actual bids received and hence the actual proceeds.

As we will see in the section \ref{sec:Exclusive-Valuation}, valuation
of this portfolio requires understanding uncertainty from numerous
angles. As the participants try to find better and improved ways to
contend with this uncertainty (Kashyap 2017), we will see that the
profitability of using this mechanism might decrease for participants
from both sides. This can lead us to believe that over time, as better
valuation methods are used by the participants, in an iterative fashion,
the profits will continue to erode. A key duty of lending desks is
the management of collateral, which can lead to the movement of securities
multiple times across many parties, exacerbating financial risk (End-note
\ref{enu:=00005BComment-Only:-Lending}). Things can go drastically
wrong even in simple environments (Sweeney \& Sweeney 1977 has a discussion
of chaos in a baby sitting monetary system with around 150 households),
hence in a complex valuation of the sort that we are dealing with
here, extreme caution should be the rule rather than the exception.
(Kashyap 2017) looks at recent empirical examples related to trading
costs where unintended consequences set in. With the above background
in mind, let us look at how we could value an exclusive portfolio.

\subsection{\label{sec:Related-Literature}Related Literature}

While the main results we develop for exclusive valuations are immediately
applicable to equity instruments, the key distinction between consumption
assets and investment assets {[}shorting generally applies only to
investments assets, whereas consumption assets cannot be directly
shorted{]} should tell us that our valuation methodologies can be
used to transfer the rights on any bundle of investment assets. (Hull
2010) has a description of consumption and investment assets, specific
to the price determination of futures and forwards; (Kashyap 2017,
End-note \ref{enu:Despite-the-several}) have a more general discussion.
The price effect on consumption assets affects the quantity bought
and consumed, whereas with investment assets (especially ones that
can be shorted), the cyclical linkage between vacillating prices and
increasing numbers of transactions is more readily apparent.

The primary focus on short selling is that activity in the shorting
market can be used to predict future security returns. There are many
studies that develop theoretical models and perform an application
of these concepts to different data-sets, both public and proprietary.
By looking at the below studies on securities lending, it becomes
clear that there is hardly any paper that considers the motivations
of the main players, the actions that arise due to these incentives
and the impact of these actions on the securities lending market and
no study at all on how to determine the value of exclusive auctions.
The below studies can also be useful for understanding the importance
of securities lending towards better portfolio performance (even if
a portfolio has no short sales) and why more research towards uncovering
the actions of the main players in the lending market can be helpful
towards this goal.

(Duffie, Garleanu \& Pedersen 2002) present a model of asset valuation
in which short-selling is achieved by searching for security lenders
and by bargaining over the terms of the lending fee. They provide
a closed-form equilibrium solution, including the dynamics of the
price, of the lending fees, and of the short interest. The price is
elevated by the prospect of future lending fees, and may, in the beginning,
be even higher than the valuation of the most optimistic agent. (Harrison
\& Kreps 1978; Morris 1996) obtain a similar result but explained
due to speculative behavior, or the right that investors hold to resell
securities, which makes them willing to pay more for it than they
would pay if they were obliged to hold it forever. (Hong \& Stein
2003) develop a theory of market crashes based on differences of opinion
among investors, with a suggestion that short-sales constraints may
play a bigger role than one might have guessed based on just the direct
transactions costs associated with shorting. 

(Diamond \& Verrecchia 1987) provide a theoretical model which implies
that the costs associated with short selling will squeeze liquidity
traders out of such order flow. This has the effect of making short
orders more informative than the population of regular sell orders.
(Allen, Morris \& Postlewaite 1993) show that even if there is a finite
number of trading opportunities, the market price of a security can
be above the present value of its future dividends, that is a bubble
can persist in the presence of asymmetric information (or agents do
not know the beliefs of other agents) with short-sales constraints.
{[}For other theoretical work on the implications of short sale constraints
for stock prices, see (Jarrow 1980) and (Scheinkman \& Xiong 2003){]}.

The standard empirical approach to testing the relation between the
shorting market and future stock returns relies on finding an appropriate
measure of short sale constraints. This measure is usually obtained
either from data on direct costs of shorting from the stock loan market,
or by employing proxies for shorting demand or shorting supply. The
idea behind looking at shorting demand is that some investors may
want to short a stock but may be impeded by constraints; if one can
measure the size of this group of investors, one can measure the extent
of overpricing or the extent of private information left out of the
market. The idea behind looking at shorting supply is that since shorting
a stock requires one first to borrow the shares, a low supply of lend-able
shares may indicate that short sale constraints are binding tightly.

(Aitken, Frino, McCorry \& Swan 1998) build on prior research by extending
the investigation of market reaction to short sales to an intraday
framework in a setting where short trades are transparent shortly
after the time of execution. Focusing on the Australian market, they
find a significantly negative abnormal return in calendar time following
short sales (initiated using both market and limit ask orders). (Bris,
Goetzmann \& Zhu 2007) analyze cross-sectional and time series information
from forty-six equity markets around the world, to consider whether
short sales restrictions affect the efficiency of the market, and
the distributional characteristics of returns to individual stocks
and market indices. They find some evidence that in markets where
short selling is either prohibited or not practiced, market returns
display significantly less negative skewness. However, at the individual
stock level, short sales restrictions appear to make no difference. 

(Boehmer, Jones \& Zhang 2008) use a panel of proprietary system order
data from the New York Stock Exchange to examine the incidence and
information content of various kinds of short sale orders. Their findings
indicate that institutional short sellers have identified and acted
on important value-relevant information that has not yet been impounded
into price. The results are strongly consistent with the emerging
consensus in financial economics that short sellers possess important
information, and their trades are important contributors to more efficient
stock prices. 

(Desai, Ramesh, Thiagarajan \& Balachandran 2002) examine stocks on
the NASDAQ and find that heavily shorted firms experience significant
negative abnormal returns after controlling for market, size, book-to-market
and momentum factors. The negative returns increase with the level
of short interest, indicating that a higher level of short interest
is a stronger bearish signals. (D’avolio 2002) describes the market
for borrowing and lending U.S. equities and provides an empirical
summary of conditions that can generate and sustain short sale constraints
(defined as legal, institutional or cost impediments to selling securities
short). (Cohen, Diether and Malloy 2007) examine the link between
the shorting market and stock prices using proprietary data from an
intermediary. They find that an increase in shorting demand leads
to negative abnormal returns. (Kolasinski, Reed \& Ringgenberg 2013)
empirically show that search frictions are related to loan fee dispersion
by examining the (Duffie, Garleanu \& Pedersen 2002) model. {[}Other
empirical studies include: (Jones \& Lamont 2002; Reed 2002; Geczy,
Musto, and Reed 2007; Mitchell, Pulvino \& Stafford 2002; Ofek \&
Richardson 2003; and Ofek, Richardson, \& Whitelaw 2003); among others.{]}

\section{\label{sec:Exclusive-Valuation}Exclusive Valuation}

\subsection{\label{subsec:Valuation-Setup}Valuation Setup}

The objective of a rational risk neutral decision maker at the intermediary
would be to maximize the profits, $P$, by utilizing the shares available
from the exclusive over the entire duration of the contract, extending
from time period, $t=0\;to\;t=T$ (Eq. \ref{eq:1}). We define all
the variables as we introduce them in the text but section \ref{sec:Appendix:-Dictionary-of}
has a complete dictionary of all the notation and symbols used in
the main results. $\upsilon,$ is the valuation of the exclusive for
the duration of the contract. A trivial result when the valuation
is zero, $\upsilon=0$ will lead to the maximum amount of profits,
but it should be clear that very low valuations will not lead to securing
the exclusive contract. Higher the valuation, higher the chances that
the exclusive becomes less elusive. Hence the goal is to obtain maximum
bounds for the valuation, above which it will not be profitable for
the intermediary.

It is worth highlighting that the decision process of the intermediary
(or the variables that he can directly influence or set) will only
include the amount of shares he can take from the exclusive, $A_{it}$,
and the additional supply of shares that can be sourced from other
beneficial owners, $O_{it}$. Here, subscript $i$ denotes the $\text{i}^{th}$
security in the portfolio and the number securities ranges from $i=1\;to\;i=n$.
The other variables are taken as exogenous: $H_{it}$, the holdings
available in the Exclusive pool; $R_{it}$, the rate on the stock
loan charged by the intermediary; $S_{it}$, the security price at
a particular time; $\beta$ is the discount factor. This assumption
is the most realistic scenario, but depending on the size of the exclusive
and internal inventory the loan rates can further be taken as variables
he can influence. What happens in practice is that there is usually
a baseline for the loan rates and a spread is added on top of it.
A deeper discussion of how loan rates are set, including the addition
of a spread component, are taken up in a separate paper (Kashyap 2016).

We get two constraints (Eq. \ref{eq:2}, \ref{eq:3}) based on the
properties of the different variables. One is that the amount of shares
for a particular security taken from the exclusive is less than the
total size of holdings in that security in the exclusive portfolio,
$A_{it}\leq H_{it}$. Clearly, $0\leq A_{it}$ is a trivial condition.
The other one is that the loan book carried by the desk, (existing
amount loaned out to external borrowers and hence can also be termed
the borrow book; though certain intermediaries reserve the words ``borrow
book ``to denote the amount they have borrowed from external lenders;
in our paper we these terms interchangeably), $B_{it}$, plus any
additional demand received based on the Locate requests $L_{it}$
and$\delta_{it}\in\left[0,1\right],$ the conversion rate of locates
into borrows must equal (to be precise, it should be greater than)
the sum of the Internal Inventory the intermediary holds based on
the positions of all trading desks within the firm, $I_{it}$, the
total holdings in the exclusive, $H_{it}$ and the additional supply
of shares, $O_{it}$. Usually securities lending desks have a certain
preferential treatment of different sources of inventory. Their primary
source is the internal inventory, since lending it out would reduce
the funding rate for their firm positions. The next source will be
any exclusive arrangements, since a fee would need to be paid irrespective
of whether these positions are used or not. The last resort is to
obtain shares from external lenders. Only when the supply from a more
preferred source is exhausted will a desk look to the next source
on its list of suppliers.

\begin{align}
P & =\underset{A_{it}}{\max}\;E_{0}\left\{ \sum_{t=0}^{T}\beta^{t}\left(\sum_{i=1}^{n}A_{it}S_{it}R_{it}\right)-\upsilon\left[\sum_{t=0}^{T}\beta^{t}\left(\sum_{i=1}^{n}H_{it}S_{it}\right)\right]\right\} \label{eq:1}
\end{align}
\begin{align}
\text{s.t.}0 & \leq A_{it}\leq H_{it}\label{eq:2}\\
 & I_{it}+H_{it}+O_{it}\leq B_{it}+\delta_{it}L_{it}\label{eq:3}
\end{align}
We can model the security prices, loan rates, the loan book, internal
inventory and exclusive holdings as GBMs. For simplicity in the numerical
results, we assume that the Weiner process governing each of these
is independent. The loan book, the internal inventory and holdings
represent number of shares, and hence are always positive making them
good candidates to be modeled as GBMs. Security prices are commonly
modeled as GBMs. In addition, our baseline models are diffusions without
mean-reversion which we can justify since an exclusive contract is
usually agreed for one to two years and the variables will not take
on excessively large values in this duration (non-negative drift rates
can grow a variable to infinity over time, but some of our variables
have negative drift rates as well as we see in the numerical results
in section \ref{sec:Data-set-Construction}). (Hull 2010) provides
an excellent account of using GBMs to model stock prices and other
time series that are always positive; See (Norstad 1999) for a discussion
of the log normal discussion.

The borrow process is highly volatile, with the the order of magnitude
of the change in the total amount of shares lent out over a few months
being multiple times of the total amount carried at any point in time.
The internal inventory can change significantly as well, though there
would be less turnover compared to the borrow process. This would
of course depend on which parts of the firm the inventory is coming
from. The holdings of the exclusive are the least volatile of the
three processes that govern shares (or at-least the intermediary would
hope so). The volatility of inventory turnover (or any supply) can
be a sign of the quality of the inventory and this can be used to
come up with a rate accordingly (the rate here is the price of the
inventory). This extension and other improvements where the loan rates
and the internal inventory can be made endogenous as opposed to the
present simplification where they are exogenous will be considered
in a subsequent paper (Kashyap 2016). 
\begin{equation}
\text{Geometric Brownian Motion }\equiv\begin{cases}
\frac{dS_{it}}{S_{it}} & =\mu_{S_{i}}dt+\sigma_{S_{i}}dW_{t}^{S_{i}}\\
\frac{dR_{it}}{R_{it}} & =\mu_{R_{i}}dt+\sigma_{R_{i}}dW_{t}^{R_{i}}\\
\frac{dB_{it}}{B_{it}} & =\mu_{B_{i}}dt+\sigma_{B_{i}}dW_{t}^{B_{i}}\\
\frac{dI_{it}}{I_{it}} & =\mu_{I_{i}}dt+\sigma_{I_{i}}dW_{t}^{I_{i}}\\
\frac{dH_{it}}{H_{it}} & =\mu_{H_{i}}dt+\sigma_{H_{i}}dW_{t}^{H_{i}}
\end{cases}\label{eq:4}
\end{equation}
\begin{align*}
\text{Geometric Brownian Motion} & \Longleftrightarrow\text{Log\;\ Normal\;\ Processes}\\
W_{t}^{X_{i}} & \Longleftrightarrow\text{Weiner\;\ Process\;\ governing}\;X_{i}^{th}\;\text{variable}.\\
E(dW_{t}^{X_{i}}dW_{t}^{X_{j}}) & =\rho_{X_{i},X_{j}}dt\;\quad=0\\
\rho_{X_{i},X_{j}} & \Longleftrightarrow\text{Correlation\;\ between}\;W_{t}^{X_{i}}\:and\:W_{t}^{X_{j}}\\
X_{i} & \in\left\{ S_{i},R_{i},B_{i},I_{i},H_{i}\right\} 
\end{align*}
The locate process is more precisely modeled as a Poisson process
since it would be reasonably accurate to consider locates as discrete
events occurring in time, that is, requests for a certain number of
shares being received in a given time interval. Given that most of
the time, the number and size of the share requests can be large,
we would need to use a high value of the arrival rate, $\lambda_{i}$.
Hence, we approximate this poison process as the absolute value of
a normal distribution with appropriate units (Cheng 1949). This introduces
a certain amount of skew, which is naturally inherent in this process.
\begin{align}
\text{Prob}\left(L_{it}\right) & =\frac{e^{-\lambda_{i}}\left(\lambda_{i}\right)^{L_{it}}}{\left(L_{it}\right)!}\label{eq:5}\\
\text{Locate\;\ Process} & \Longleftrightarrow\text{Poission\;\ Process\;\ with\;\ Arrival\;\ Rate},\;\lambda_{i}\\
\text{Alternately},\;L_{it} & \sim\left|N\left(\mu_{L_{i}},\sigma_{L_{i}}^{2}\right)\right|,\;\text{Absolute\;\ Normal\;\ Distribution}
\end{align}
With the above variables and their properties, we consider ways in
which we can simplify the system and obtain solutions that can aid
in putting a numerical value on the portfolio of securities. In a
complex system (\ref{eq:1}, \ref{eq:2}, \ref{eq:3}, \ref{eq:4},
\ref{eq:5} and \ref{eq:8}), deriving equations can be a daunting
exercise, and not to mention, of limited practical validity. Hence,
to supplements equations, we will employ simplifications that establish
a few inequalities governing this system. Pondering on the sources
of uncertainty and the tools we have to capture it, might lead us
to believe that, either, the level of our mathematical knowledge is
not advanced enough, or, we are using the wrong methods. The dichotomy
between logic and randomness is a topic for another time. 

\subsection{\label{subsec:Benchmark-Valuation-or}Benchmark Valuation or Zero
Profits Upper Bounds}

To obtain an upper bound for the profits, we note that if the cash
flows received from the loans made using the shares from the exclusive
exactly balance out the payments to be made to the exclusive portfolio
owner, we would have zero profits. If this is the criteria under which
we obtain zero profits, then it becomes the maximum value one would
be willing to pay for the exclusive contract; any actual valuation
for the exclusive should be less than this zero profit valuation:
$\upsilon^{actual}=\upsilon\leq\upsilon^{zero}$. Here, we need to
remember that we only use exclusives when the total demand faced by
the lending desk is higher than the internal inventory available to
the desk. This can be written as, $\text{when}\;B_{it}+\delta_{i}L_{it}\leq I_{it}\;\text{then}\;A_{it}=0$.
Combining it with the constraint (Eq. \ref{eq:2}, $A_{it}\leq H_{it}$),
that is we cannot obtain shares greater than the total size of the
holdings in the exclusive, gives us a result captured in the below
proposition.
\begin{prop}
\label{1-The-zero-profits}The zero profits upper bound for the valuation
is given by 
\begin{equation}
\upsilon^{actual}=\upsilon\leq\upsilon^{zero}=E_{0}\left\{ \frac{\sum_{t=0}^{T}\sum_{i=1}^{n}\beta^{t}\min\left[H_{it},\max\left(B_{it}+\delta_{i}L_{it}-I_{it},0\right)\right]S_{it}R_{it}}{\sum_{t=0}^{T}\beta^{t}\left(\sum_{i=1}^{n}H_{it}S_{it}\right)}\right\} \label{eq:8}
\end{equation}
\end{prop}
\begin{proof}
See appendix \ref{subsec:Proof-of-Proposition-Zero-Profits}.
\end{proof}
An immediate takeaway from this expression, which has immense intuitive
appeal, is that the higher the valuation lesser will be the extent
of overlap between the internal inventory and the exclusive holding.
Checking this historical overlap between internal inventory and exclusive
holdings can be an excellent complement to the valuation expressions
we derive.

A standard theoretical approach to solving (Eq. \ref{eq:8}) or obtaining
a closed forum solution, is presently unknown to the best of our knowledge,
given the number of GBMs the system incorporates. An alternate approach
would be to estimate the parameters of all the random variables from
historical data and run simulations that would provide the required
exclusive valuation. It is worth keeping in mind that the intermediary
firm or the beneficial owner will have access to a historical time
series of some of the variables and hence can estimate the actual
process for the various variables. Though either party will not know
the time series of all the variables with certainty and hence would
need to substitute the unknown variables purely with a simulation
based process similar to what we have used in section \ref{sec:Data-set-Construction}.
(Campbell, Lo, MacKinlay \& Whitelaw 1998; Lai \& Xing 2008; Cochrane
2009) are handy resources on using maximum likelihood estimation (MLE)
and generalized method of moments (GMM) for parameter estimation.
A simplification is to assume that the variables are independent.
While a realistic calculation might show that these variables are
correlated, such a simplification provides an excellent benchmark
for our valuation exercise; (Gujarati 1995; Hamilton 1994) discuss
time series simplifications and the need for parsimonious models .
A backward induction based computer program, which simulates the randomness
component of the variables involved, can calculate the value of the
exclusive based on the above expression (Eq. \ref{eq:8}).

As a further simplification in this upper limit (Eq. \ref{eq:8})
for the valuation, we set $\beta=1$. We then get the $\upsilon^{beta}$
valuation below, which is used as the foundation to derive other alternative
expressions in section (\ref{subsec:Valuation-with-Transaction},
\ref{subsec:Other-Conservative-Valuation} and \ref{subsec:Historical-Valuations}),
\begin{equation}
\upsilon^{beta}=E_{0}\left\{ \frac{\sum_{t=0}^{T}\sum_{i=1}^{n}\min\left[H_{it},\max\left(B_{it}+\delta_{i}L_{it}-I_{it},0\right)\right]S_{it}R_{it}}{\sum_{t=0}^{T}\left(\sum_{i=1}^{n}H_{it}S_{it}\right)}\right\} \label{eq:9}
\end{equation}

\subsection{\label{subsec:Valuation-with-Transaction}Valuation with Transaction
Costs}

It is not uncommon to have a transaction cost, $TC$, when securities
are taken out or put back into an exclusive portfolio. Hence, it is
useful to have a valuation expression, $\upsilon^{transaction}$,
after incorporating transaction costs.
\begin{prop}
\label{2-The-valuation-expression}The valuation expression that captures
transaction costs is given by
\begin{equation}
\upsilon^{transaction}=E_{0}\left\{ \frac{\left(\sum_{t=0}^{T}\sum_{i=1}^{n}\min\left[H_{it},\max\left(B_{it}+\delta_{i}L_{it}-I_{it},0\right)\right]S_{it}R_{it}\right)-\left(TC\right)}{\left(\sum_{t=0}^{T}\sum_{i=1}^{n}H_{it}S_{it}\right)}\right\} \label{eq:10}
\end{equation}
Here,
\begin{eqnarray*}
\text{Transaction\;\ Costs\;}\equiv TC & = & E_{0}\left\{ \sum_{i=1}^{n}c\left\{ \frac{\max\left(B_{i0}+\delta_{i}L_{i0}-I_{i0},0\right)}{\left(B_{i0}+\delta_{i}L_{i0}-I_{i0}\right)}\right\} \right.\\
 & + & \sum_{t=1}^{T}\sum_{i=1}^{n}\frac{c}{2}\left[\left|\left\{ \frac{\max\left(B_{it}+\delta_{i}L_{it}-I_{it},0\right)}{\left(B_{it}+\delta_{i}L_{it}-I_{it}\right)}\right.\right.\right.\\
 & - & \left.\frac{\max\left(B_{it-1}+\delta_{i}L_{it-1}-I_{it-1},0\right)}{\left(B_{it-1}+\delta_{i}L_{it-1}-I_{it-1}\right)}\right\} \\
 & - & \left\{ \frac{\max\left(I_{it}-B_{it}-\delta_{i}L_{it},0\right)}{\left(I_{it}-B_{it}-\delta_{i}L_{it}\right)}\right.\\
 & - & \left.\left.\left.\left.\frac{\max\left(I_{it-1}-B_{it-1}-\delta_{i}L_{it-1},0\right)}{\left(I_{it-1}-B_{it-1}-\delta_{i}L_{it-1}\right)}\right\} \right|\right]\right\} 
\end{eqnarray*}
\end{prop}
\begin{proof}
See appendix \ref{subsec:Proof-of-Proposition-2-Transaction-Costs}.
\end{proof}
It is trivial to see that,
\begin{equation}
\upsilon^{transaction}\leq\upsilon=\upsilon^{actual}\leq\upsilon^{zero}\label{eq:11}
\end{equation}
In a similar vein, we can also arrive at an expression for transaction
costs when the charges to Take and Give are different. We don't derive
that here, since that is usually a rarity. Alternately, (Erdos and
Hunt 1953) derive results regarding the change of signs of sums of
random variables which can provide approximations for transaction
costs.

\subsection{\label{subsec:Other-Conservative-Valuation}Other Conservative Valuation
Inequalities to Supplement Equations}

We now provide various methods to come up with more conservative estimates
of the valuation. We call these conservative valuations because they
underestimate the potential benefit or profits such a valuation would
bring. As we can see from (Eq. \ref{eq:9}), there are two variables
that can be useful towards this end. First, we can set $\delta_{i}=0$
or when excess demand is zero (other alternatives are lower values
of $\delta_{i}$), giving us a conservative valuation, $\upsilon^{conservative}$,with
zero excess demand,
\begin{equation}
\Rightarrow\upsilon^{conservative}=E_{0}\left\{ \frac{\sum_{t=0}^{T}\sum_{i=1}^{n}\min\left[H_{it},\max\left(B_{it}-I_{it},0\right)\right]S_{it}R_{it}}{\sum_{t=0}^{T}\left(\sum_{i=1}^{n}H_{it}S_{it}\right)}\right\} \label{eq:12}
\end{equation}
Instead of using the rates at which the desk makes loans to borrowers,
we can use the rate at which it finds supply from other beneficial
owners. Sometimes, where no other supply is available a theoretical
rate is used by lending desks. The different possible variations here
would depend on the different types of rates (possibly due to different
levels of spread) a lending desk would use on a daily basis and also
store historically. We show two variations one which has a lower value
of the loan rates, $\upsilon^{beta\;alternate}$, and another combines
zero excess demand with the lower value of the loan rates , $\upsilon^{alternate}$,
\begin{equation}
\Rightarrow\upsilon^{beta\;alternate}=E_{0}\left\{ \frac{\sum_{t=0}^{T}\sum_{i=1}^{n}\min\left[H_{it},\max\left(B_{it}+\delta_{i}L_{it}-I_{it},0\right)\right]S_{it}Q_{it}}{\sum_{t=0}^{T}\left(\sum_{i=1}^{n}H_{it}S_{it}\right)}\right\} \label{eq:13}
\end{equation}
 
\begin{equation}
\Rightarrow\upsilon^{alternate}=E_{0}\left\{ \frac{\sum_{t=0}^{T}\sum_{i=1}^{n}\min\left[H_{it},\max\left(B_{it}-I_{it},0\right)\right]S_{it}Q_{it}}{\sum_{t=0}^{T}\left(\sum_{i=1}^{n}H_{it}S_{it}\right)}\right\} \label{eq:14}
\end{equation}
This gives the following pecking order of valuations for the exclusive.
\begin{equation}
\upsilon^{beta}\geq\upsilon^{conservative}\geq\upsilon^{alternate}\label{eq:15}
\end{equation}
\begin{equation}
\upsilon^{beta}\geq\upsilon^{beta\;alternate}\geq\upsilon^{alternate}\label{eq:16}
\end{equation}
The intermediary can decide on their level of aggressiveness and choose
which of the valuations they wants to use, depending on how many exclusives
they already have, the extent of overlap with their internal inventory,
the number of special names in the exclusive portfolio and the volatility
of the time series of daily profits from the exclusive. Such a tiered
approach is found to be more practical rather than having an exact
valuation since there are too many sources of uncertainty and the
noise or the variance of any exact valuation number would tend to
be high.

\subsection{\label{subsec:Historical-Valuations}Historical Valuations}

Given the complexity and the number of variables to be estimated (Eqs:
\ref{eq:1}, \ref{eq:2}, \ref{eq:3}, \ref{eq:4}, \ref{eq:5} and
\ref{eq:8}), a simple heuristic would be utilize the historical time
series of each of the variables and then use that as a possible guide
to the calculation of the exclusive value. The theoretical justification
for using the actual historical series directly to value the exclusive
without first resorting to parameter estimations and then running
simulations using the estimated distribution parameters is that there
are errors introduced during the estimation which are then compounded
while doing the simulations (Kashyap 2017). The pecking order shown
above (Eqs: \ref{eq:11}, \ref{eq:15} and \ref{eq:16}) can be arrived
at using the historical time series as well. Using this, we can also
arrive at the time series of the daily profits that would accrue to
the intermediary. The volatility of the daily value of the exclusive
can be suggestive in terms of how aggressive one should be in picking
one of the valuation tiers. Any of the historical valuations $\upsilon^{historical}$
can be represented as below, where $-T_{s}$ and $-T_{e}$ denote
the start and end of the time series such that $T_{s}\geq T_{e}$,
\begin{equation}
\Rightarrow\upsilon^{historical}=\left\{ \frac{\sum_{t=-T_{s}}^{-T_{e}}\sum_{i=1}^{n}\min\left[H_{it},\max\left(B_{it}+\delta_{i}L_{it}-I_{it},0\right)\right]S_{it}R_{it}}{\sum_{t=-T_{s}}^{-T_{e}}\left(\sum_{i=1}^{n}H_{it}S_{it}\right)}\right\} \label{eq:17}
\end{equation}
New valuation time series can be created by adding transaction costs
or alternate rates, or other combinations. Armed with this set of
valuations, $\left\{ \upsilon^{zero},\upsilon^{beta},\upsilon^{beta\;alternate},\upsilon^{transaction},\right.$
$\left.\upsilon^{conservative},\upsilon^{alternate},\upsilon^{historical}\right\} $,
the bidder can combine them using the method we shown in the next
sub-section (\ref{subsec:Variance-Weighted-Combined}). 

\subsection{\label{subsec:Variance-Weighted-Combined}Variance Weighted Combined
Valuation}

\textbf{\textit{A valuation is nothing but a forecast of future value
or benefits that we hope to derive from any object. }}There is a huge
amount of literature on combining forecasts starting with the seminal
paper by (Bates \& Granger 1969). For different methods of combining
forecasts including recent developments and surveys on the topic see:
(Granger \& Ramanathan 1984; Granger 1989; Clemen 1989; Timmermann
2006; Hsiao \& Wan 2014; Conflitti, De Mol \& Giannone 2015; Chan
\& Pauwels 2018). (Smith \& Wallis 2009) has a formal explanation
of the forecast combination puzzle, that simple combinations of point
forecasts are repeatedly found to outperform sophisticated weighted
combinations in empirical applications. The minimum variance weights
for the $i^{th}$valuation $\upsilon_{i}$ among a set of $k$ valuations
are given by $w_{i}=\frac{1}{\sigma_{i}^{2}}/\left(\underset{j=1}{\overset{k}{\sum}}\frac{1}{\sigma_{j}^{2}}\right)$.
We give a proof that this is the minimum variance weighting as the
number of valuations increases in appendix \ref{subsec:Minimum-Variance-Weighting}.

\textbf{\textit{In theorem \ref{Th-1-When-each-of} we show an alternative
weighting scheme that combines the valuations using the variance of
the individual valuation time series. We derive the conditions under
which it converges to the true valuation faster than the minimum variance
combination. Hence, the results in this section are applicable to
problems where multiple forecasts are obtained and we need a technique
to weight the forecasts to come up with a single value. This shows
that this weighting methodology could have wide usage outside of the
securities lending or even the financial industry.}}

We argue that under certain conditions of finite variance and finite
valuation of each individual time series we get closer to the true
valuation as the number of individual time series considered in the
aggregation gets larger. For simplicity of notation, in this section
we let each individual time series be represented by $\upsilon_{i},\;i\in\left\{ 1,2,...,k\right\} $
with corresponding variances $\sigma_{i}^{2}$ and the true valuation
by $\upsilon$.
\begin{thm}
\label{Th-1-When-each-of}If each of the valuations and the variances
of the valuations are finite, $\left\{ \upsilon_{i}<\infty;\sigma_{i}^{2}<\infty\right\} $;
the covariances are zero, $cov\left(\upsilon_{i},\upsilon_{j}\right)=0$;
no single one dominates the sum, expressed as,
\begin{eqnarray*}
\left[\underset{k\rightarrow\infty}{\lim}\frac{\max\left(\sigma_{i}^{2}\right)}{\underset{i=1}{\overset{k}{\sum}}\;\sigma_{i}^{2}}\rightarrow0\;;\;\underset{k\rightarrow\infty}{\lim}\frac{\max\left(\sigma_{i}^{2}\upsilon_{i}\right)}{\underset{i=1}{\overset{k}{\sum}}\;\sigma_{i}^{2}\upsilon_{i}}\rightarrow0\right]
\end{eqnarray*}
and the following combination of the valuations and variances is uniformly
bounded, that is for any real number $M_{1}$,
\[
\text{}\frac{\left(\sum_{i=1}^{k}\sigma_{i}^{2}v_{i}\right)}{(\sum_{i=1}^{k}\sigma_{i}^{2})}\leq M_{1}
\]
Then when each of the individual valuations are weighted using the
scheme shown below (sum of the variance of all other valuations divided
by the total variance times the number of valuations) the expression
below asymptotically converges in probability ($\overset{p}{\longrightarrow}$)
to the true valuation.
\[
E\left[\underset{k\rightarrow\infty}{\lim}\frac{1}{\left(k\right)}\frac{\underset{i=1}{\overset{k}{\sum}}\;\underset{j\neq i}{\overset{k}{\sum}}\;\sigma_{j}^{2}\upsilon_{i}}{\underset{i=1}{\overset{k}{\sum}}\sigma_{i}^{2}}\right]\quad\overset{p}{\longrightarrow}\quad E\left[\upsilon\right]\quad\text{ When the valuations have the same mean.}
\]
When the valuations have different expected values we have the result,
\[
E\left[\underset{k\rightarrow\infty}{\lim}\frac{1}{\left(k\right)}\frac{\underset{i=1}{\overset{k}{\sum}}\;\underset{j\neq i}{\overset{k}{\sum}}\;\sigma_{j}^{2}\upsilon_{i}}{\underset{i=1}{\overset{k}{\sum}}\sigma_{i}^{2}}\right]\quad\overset{p}{\longrightarrow}\quad E\left[\bar{\upsilon}_{k}\right]\quad Here,\ensuremath{\bar{\upsilon}_{k}=\frac{1}{\left(k\right)}\left\{ \underset{i=1}{\overset{k}{\sum}}\left(\upsilon_{i}\right)\right\} }\text{ }
\]
This alternative weighting scheme converges faster to the true valuation
than the minimum variance weighting scheme when the following conditions
hold,
\[
\left(\frac{\sigma_{k}^{2}}{\sigma_{k+1}^{2}}\right)>1\;;\;\left(\frac{\upsilon_{k+1}}{\upsilon_{k}}\right)<1\;;\;\left(\frac{\upsilon_{k+1}}{\upsilon_{k}}\right)\left(\frac{\sigma_{k}^{2}}{\sigma_{k+1}^{2}}\right)<1
\]
\end{thm}
\begin{proof}
See appendix \ref{subsec:Proof-of-Theorem-Variance-Weighted} for
the proofs, including expressions for the variance of the aggregated
valuations (both the alternative weighting and minimum variance schemes)
and the corresponding asymptotic result showing that the variance
of the combined valuation as $k\rightarrow\infty$ is zero.
\end{proof}
Similar to the minimum variance weighting, our alternative weighting
scheme has an intuitive and practical appeal since the time series
with a higher variance is set a lower weight in the combined valuation.
But if we have many valuations and we find that the conditions for
faster convergence are satisfied our alternative scheme becomes more
desirable. This means the more expressions we are able to derive for
the valuations and combine them, the better will be our estimation.
Of course, it becomes important to ensure that we do not have redundant
valuation expressions that are just multiples of one other; but valuations
that would be good candidates to vary and create newer time series
are the ones that differ by capturing the different possible variations
in any of the variables that can affect the valuation outcome. This
result has a certain theoretical significance since it shows that
when any object has multiple valuations, where each valuation might
arise due to slightly different assumptions; a combination using our
technique gets closer to the true valuation faster.

Alternately, the intermediary can subjectively select a particular
valuation to suit the institutional setup or relevance depending on
the preferences at the intermediary (whether their internal inventory
tends to be large or they are looking to have many exclusives in place
expecting greater future demand and so on). Either way, once a final
valuation has been obtained the next step is to prepare for an auction
process and pick a strategy that will shade the value to suit the
mechanics of different auction situations.

\section{\label{sec:Auction-Strategy}Auction Strategy}

Once we have the valuation from the previous section (\ref{sec:Exclusive-Valuation}),
it is important to look at different auction formats and the specifics
of how an intermediary would tailor bids to adapt to the particular
auction setting. From the perspective of the owner of the exclusive
portfolio, he would use the valuation and the auction setting to understand
the potential revenue opportunity. We consider a few variations in
the first price sealed bid auction mechanism. \textbf{\textit{The
key auction theory results we use, including the proofs for some of
the extensions of core auction theory results to real life applications,
are from (Kashyap 2018). }}The results we discuss below are not new,
but we provide them for completeness and also because they are important
for the numerical calculations in section \ref{sec:Model-Testing-Results}. 

The literature on Auction Theory is vast and deep. The following standard
and detailed texts on this topic might aid the interested reader:
(Klemperer 2004; Krishna 2009; Menezes \& Monteiro 2005; and Milgrom
2004). Additional references are (Laffont, Ossard \& Vuong 1995; Milgrom
\& Weber 1982); for using numerical techniques (Miranda \& Fackler
2002) or approximations to the error function (Chiani, Dardari and
Simon 2003) . (Ortega-Reichert 1967; and Harstad, Kagel \& Levin 1990)
derive the expression when there is uncertainty about the number of
bidders. (Levin \& Ozdenoren 2004; and Dyer, Kagel \& Levin 1989)
are other useful references. (Lebrun 1999) derives conditions for
the existence of an asymmetric equilibrium with more than two bidders.

A bidding strategy is sensitive to assumed distributions of both the
valuations and the number of bidders. This is seen from the expression
for the bid strategy in (Lemma \ref{Lemma-The-symmetric-equilibrium}).
As a benchmark bidding case, it is illustrative to assume that all
bidders know their valuations and only theirs and they believe that
the values of the others are independently distributed according to
the general distribution $F$. $x_{i}$ is the valuation of bidder
$i$. This is a realization of the random variable $X_{i}$ which
bidder $i$ and only bidder $i$ knows for sure. $x_{i}\sim F\left[0,\omega\right]$,
$x_{i}$ is symmetric and independently distributed according to the
distribution $F$ over the interval $\left[0,\omega\right]$. $F,$
is increasing and has full support, which is the non-negative real
line $\left[0,\infty\right]$, hence in this formulation we can have
$\omega=\infty$. $f=F',$ is the continuous density function of $F$.
$M,$ is the total number of bidders. When there is no confusion about
which specific bidder we are referring to, we drop the subscripts
such as in the valuation $x$. $Y_{1}\equiv Y_{1}^{M-1}$ , is the
random variable that denotes the highest value, say for bidder 1,
among the $M-1$ other bidders. $Y_{1},$ is the highest order statistic
of $X_{2},X_{3},...,X_{M}$. $G,$ is the distribution function of
$Y_{1}$. That is, $\forall y,\;G(y)=\left[F(y)\right]^{M-1}$and
$g=G',$ is the continuous density function of $G$ or $Y_{1}$. $m\left(x\right),$
is the expected payment of a bidder with value $x$. $\beta_{i}:\left[0,\omega\right]\rightarrow\Re_{+}$
is an increasing function that gives the strategy for bidder $i$.
We let $\beta_{i}\left(x_{i}\right)=b_{i}$. We must have $\beta_{i}\left(0\right)=0$.
$\beta:\left[0,\omega\right]\rightarrow\Re_{+}$ is the strategy of
all the bidders in a symmetric equilibrium. We let $\beta\left(x\right)=b,\;x$
is the valuation of any bidder. We also have $b\leq\beta\left(x\right)\;\text{and}\;\beta\left(0\right)=0$.
\begin{lem}
\label{Lemma-The-symmetric-equilibrium}The symmetric equilibrium
bidding strategy for a bidder, the expected payment of a bidder and
the expected revenue of a seller are given by

Equilibrium Bid Function is,
\begin{eqnarray*}
\beta\left(x\right) & = & \left[x-\int_{0}^{x}\left[\frac{F\left(y\right)}{F\left(x\right)}\right]^{M-1}dy\right]
\end{eqnarray*}
Expected ex ante payment of a particular bidder is,
\begin{eqnarray*}
E\left[m\left(x\right)\right] & = & \int_{0}^{\omega}y\left[1-F\left(y\right)\right]g\left(y\right)dy
\end{eqnarray*}

Expected revenue, $R_{s}$, to the seller is
\[
E\left[R_{s}\right]=ME\left[m\left(x\right)\right]
\]
\end{lem}
We consider two distributions for shading the valuation to formulate
a bidding strategy: Uniform and Log-normal. The two distribution types
we discuss can shed light on the other types of distributions in which
only positive observations are allowed. The uniform distribution is
well uniform and hence is ideal when the valuations (or sometimes
even the number of bidders) are expected to fall equally on a finite
number of possibilities (Corollary \ref{Corr-1-The-symmetric-equilibrium}).
This serves as one extreme to the sort of distribution we can expect
in real life.
\begin{cor}
\label{Corr-1-The-symmetric-equilibrium}The symmetric equilibrium
bidding strategy when the valuations are distributed uniformly is
given by
\begin{eqnarray*}
\beta\left(x\right) & = & \left(\frac{M-1}{M}\right)x
\end{eqnarray*}
Here, $x_{i}\sim U\left[0,\omega\right]$ since we are considering
the uniform distribution.
\end{cor}
The other case is a log-normal distribution which centers around a
value and the chance of observing values further away from this central
value become smaller. Asset prices are generally modeled as log-normal,
so financial applications, including an exclusive valuation would
benefit from this extension. The absence of a closed form solution
for the log-normal distribution forces us to develop a rough theoretical
approximation (Corollary \ref{Corr-2-The-symmetric-equilibrium})
and improve upon that significantly using non-linear regressions (Remark
\ref{A-better-approximation}; Eq. \ref{eq:Power}). This works well
for our particular application, since the valuations are generally
small, of the order of a few basis points. A detailed discussion of
the the log-normal approximation, including the accuracy of the regression
results, suggested values for the regression coefficients and the
sensitivity of the bid strategy to the valuation, the parameters of
the valuation distribution and the number of bidders is provided in
(Kashyap 2018).
\begin{cor}
\label{Corr-2-The-symmetric-equilibrium}The symmetric equilibrium
bidding strategy when the valuations are small, of the order less
than one, and distributed log-normally, can be roughly approximated
as
\begin{eqnarray*}
\beta\left(x\right) & = & \left[x-\frac{\int_{0}^{x}\left[\Phi\left(\frac{\ln y-\mu}{\sigma}\right)\right]^{M-1}dy}{\left[\Phi\left(\frac{\ln x-\mu}{\sigma}\right)\right]^{M-1}}\right]\\
 & \approx & \frac{x}{2}
\end{eqnarray*}
Here, $\Phi(u)=\frac{1}{\sqrt{2\pi}}\int_{-\infty}^{u}e^{-t^{2}/2}dt$
, is the standard normal cumulative distribution and $X=e^{W}$where,
$W\sim N\left(\mu,\sigma\right)$. $x_{i}\sim LN\left[0,\omega\right]$
since we are considering the log-normal distribution.
\end{cor}
When the number of bidders are large the above expression for the
uniform distribution (Corollary \ref{Corr-1-The-symmetric-equilibrium})
does not depend on the number of bidders, that is $\underset{M\rightarrow\infty}{\lim}\left(\frac{M-1}{M}\right)x=x$.
Comparing the bidding strategy with respect to the distribution of
valuations in the two cases, the uniform distribution when the number
of bidders are large and the log-normal distribution theoretical approximation
(Corollary \ref{Corr-2-The-symmetric-equilibrium}) we see that: 1)
both do not depend significantly on the number of bidders and 2) the
bid is larger with a uniform distribution.
\begin{rem}
\label{A-better-approximation}A better approximation for the log-normal
distribution can be obtained using non-linear regression to find the
constant, $C$, and the power coefficients, $a_{1}$,$a_{2}$,$a_{3}$
and $a_{4}$ in the expression below,
\begin{equation}
\beta\left(x\right)=Cx^{a_{1}}\mu^{a_{2}}\sigma^{a_{3}}M^{a_{4}}\label{eq:Power}
\end{equation}
\end{rem}
In many auction settings, the auction seller can set a minimum bid
to ensure that he is guaranteed a minimum amount of revenue. This
minimum bid is known as the reserve price. Our valuation techniques
can help auction sellers come up with a reserve price. Clearly, if
the reserve price is too high, many potential bidders will shy away
from participating in the auction. But setting a reserve price, ensures
that the bid strategies need to be higher to ensure successfully winning
the auction (Lemma \ref{Lemma-2-The-symmetric-equilibrium}).
\begin{lem}
\label{Lemma-2-The-symmetric-equilibrium}The symmetric equilibrium
bidding strategy when the valuation is greater than the reserve price,
$\left(r>0\right)$, of the seller, $x\geq r$, is

For a general distribution, 
\begin{eqnarray*}
\beta\left(x\right) & = & r\frac{G\left(r\right)}{G\left(x\right)}+\frac{1}{G\left(x\right)}\int_{r}^{x}yg\left(y\right)dy\\
\text{Alternately }\beta\left(x\right) & = & x-\int_{r}^{x}\frac{G\left(y\right)}{G\left(x\right)}dy
\end{eqnarray*}
\end{lem}
When the valuations are distributed uniformly, the bid strategy with
a reserve price is given in (Corollary \ref{Corr-3-The-symmetric-equilibrium}),
\begin{cor}
\label{Corr-3-The-symmetric-equilibrium}The symmetric equilibrium
bidding strategy when the valuation is greater than the reserve price
of the seller, $x\geq r$, and valuations are from an uniform distribution,
\[
\beta\left(x\right)=\frac{r^{M}}{x^{M-1}}\left(\frac{M+1}{M}\right)+x\left(\frac{M-1}{M}\right)
\]
\end{cor}
When the valuations are distributed log-normally, the bid strategy
with a reserve price is given in (Corollary \ref{Corr-4-The-symmetric-equilibrium}),
\begin{cor}
\label{Corr-4-The-symmetric-equilibrium}The symmetric equilibrium
bidding strategy when the valuation is greater than the reserve price
of the seller, $x\geq r$, and valuations are from a log normal distribution,
\[
\beta\left(x\right)=x\left[\frac{h'\left(r\right)\left(x-r\right)}{h\left(x\right)}+\frac{r}{x}\frac{h\left(r\right)}{h\left(x\right)}\right]
\]
\[
\text{Here, }h'\left(r\right)=\left(M-1\right)\left[\int_{-\infty}^{\left(\frac{lnr-\mu}{\sigma}\right)}e^{-t^{2}/2}dt\right]^{M-2}\left\{ \frac{e^{-\left(\frac{lnr-\mu}{\sigma}\right)^{2}/2}}{r\sigma}\right\} 
\]
\end{cor}
The following result can aid auction sellers in finding an optimal
reserve price (Lemma \ref{Lemma-3-The-optimal-reserve}),
\begin{lem}
\label{Lemma-3-The-optimal-reserve}The optimal reserve price for
the seller, $r^{*}$ must satisfy the following expression, 
\[
x_{s}=r^{*}-\frac{\left[1-F\left(r^{*}\right)\right]}{f\left(r^{*}\right)}
\]
Here, seller has a valuation, $x_{s}\in\left[0,\omega\right)$
\end{lem}
When there is uncertainty about how many interested bidders there
are, we denote the potential set of bidders as $\mathcal{M}=\left\{ 1,2,\ldots,M\right\} $.
$\mathcal{A}\subseteq\mathcal{M}$ is the set of actual bidders. All
potential bidders draw their valuations independently distributed
according to the general distribution $F$. Also, $p_{l}$ is the
probability that any participating bidder, $i\in\mathcal{A},$ is
facing $l$ other bidders or the probability that he assigns to the
event that he is facing $l$ other bidders. This implies that there
is a total of $l+1$ bidders, $l\in\left\{ 1,2,\ldots,M-1\right\} $.
$G^{l}\left(x\right)=\left[F\left(x\right)\right]^{l}$ is the probability
of the event that the highest of $l$ values drawn from the symmetric
distribution $F$ is less than $x$, his valuation and the bidder
wins in this case. $\beta^{l}\left(x\right)$ is the equilibrium bidding
strategy when there are a total of exactly $l+1$ bidders, known with
certainty. The overall probability that the bidder will win when he
bids $\beta^{M}\left(x\right)$ is 
\begin{equation}
G\left(x\right)=\sum_{l=0}^{M-1}p_{l}G^{l}\left(x\right)\label{eq:19}
\end{equation}

\begin{figure}[H]
\includegraphics[width=8cm]{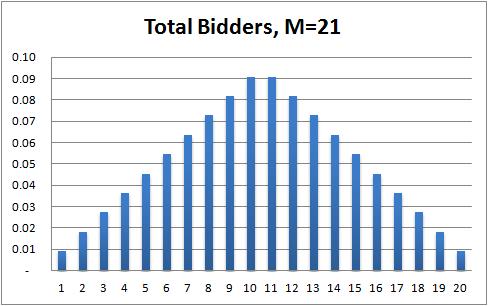}\includegraphics[width=8cm]{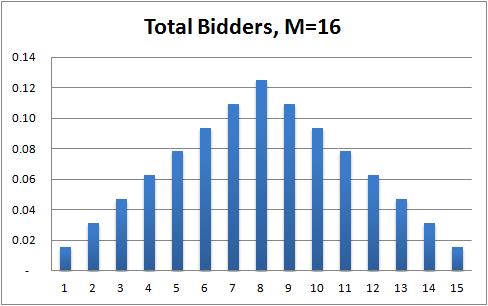}

\caption{\label{fig:Variable-Bidders-Symmetric}Variable Bidders Symmetric
Discrete Probability Distributions}
\end{figure}

Hence the equilibrium bid for an actual bidder when he is unsure about
the number of rivals he faces is a weighted average of the equilibrium
bids in an auction when the number of bidders is known to all. (McAfee
\& McMillan 1987b) is one of the most well known and early generalizations
to allow the number of bidders to be stochastic.
\begin{lem}
\label{The-equilibrium-strategy}The equilibrium strategy when there
is uncertainty about the number of bidders is given by
\begin{eqnarray*}
\beta^{M}\left(x\right) & = & \sum_{l=0}^{M-1}\frac{p_{l}G^{l}\left(x\right)}{G\left(x\right)}\beta^{l}\left(x\right)
\end{eqnarray*}
\end{lem}
When bidding for an exclusive, an intermediary, will expect most of
the other major players to be bidding as well. Invariably, there will
be some drop outs, depending on their recent exclusive bidding activity
and some smaller players will show up based on the composition of
the portfolio being auctioned. It is a reasonable assumption that
all of the bidders hold similar beliefs about the distribution of
the number of players. Hence, for the numerical results, we construct
a symmetric discrete distributions of the sort shown in (Figure \ref{fig:Variable-Bidders-Symmetric}).
This formulations of a positive symmetric discrete distribution is
likely to be followed by the total number of auction participants,
and we incorporate this into auction theory results. We show that
such a distribution satisfies all the properties of a probability
distribution function as part of the proof for Lemma \ref{The-formula-for}
(Kashyap 2018). It is to be noted that this symmetric discrete distribution
comes under the family of triangular distributions (End-note \ref{Triangular-Distribution}).
We can easily come up with variations that can provide discrete asymmetric
probabilities. For simplicity, we use the uniform distribution for
the valuations and set $\omega=1$. The below result follows from
a bidding strategy that incorporates the use of the discrete symmetric
distribution.
\begin{lem}
\label{The-formula-for}The bidding strategy and the formula for the
probability of facing any particular total number of bidders under
a symmetric discrete distribution would be given by,
\[
\beta\left(x\right)=\sum_{l=0}^{M-1}\left(\frac{p_{l}x^{l}}{\sum_{k=0}^{M-1}p_{k}x^{k}}\right)\left(\frac{l}{l+1}\right)x
\]
\begin{eqnarray*}
p_{l} & = & \begin{cases}
l\Delta_{p} & ,\;\text{if}\;\;l\leq\frac{\left(M-1\right)}{2}\\
\left(M-l\right)\Delta_{p} & ,\;\text{if}\;\;l>\frac{\left(M-1\right)}{2}
\end{cases}
\end{eqnarray*}
\[
\Delta_{p}=\left\{ \left\lfloor \frac{M^{2}}{4}\right\rfloor \right\} ^{-1}
\]
We note that $\Delta_{p}$ can also be written as,
\[
\Delta_{p}=\frac{1}{\left\{ \left\lfloor \frac{\left(M-1\right)}{2}\right\rfloor \left\{ \left\lfloor \frac{\left(M-1\right)}{2}\right\rfloor +1\right\} +\left[\left\{ \left(\frac{\left(M-1\right)}{2}\bmod1\right)+\frac{\left(M-1\right)}{2}\right\} \left\{ 2\left(\frac{\left(M-1\right)}{2}\bmod1\right)\right\} \right]\right\} }
\]
\begin{align*}
\left\lfloor \frac{\left(M-1\right)}{2}\right\rfloor \text{ is the integer floor function, that is, it rounds any number down to the nearest integer. }\\
A\bmod B\text{ is the modulo operator, that is, it gives the remainer when }A\text{ is divided by }B.\\
\text{When }A\text{ is a fraction less than one and }B\text{ is one, the result is the fraction itself.}
\end{align*}
\end{lem}

This discrete distribution can also be a possibility for the valuations
themselves, since the set of prices of assets or valuations can be
from a finite set. But given the distribution, developing a bidding
strategy based on this discussion is trivial and hence is not explicitly
considered here. Lastly, the case of interdependent valuations are
to be highly expected in real life; but practical extensions for this
case are near absent both in the literature and in practice. In appendix
\ref{sec:Auction-Theory-Results}, we provide extensions when the
valuations of the bidders are interdependent and incorporate the corresponding
results into a final combined realistic setting. We also provide additional
results when bidders hold asymmetric beliefs in appendix \ref{sec:Auction-Theory-Results}.
These results can be useful extensions to aid the profit maximization
goals of exclusive auction participants depending on the assumptions
they wish to make regarding their environment and can be useful for
bidders and auctions sellers during the wholesale procurement of other
financial instruments.

\section{\label{sec:Numerical-Results}Numerical Results}

\subsection{\label{sec:Data-set-Construction}Data-set Construction}

\begin{figure}[H]
\includegraphics[width=6cm]{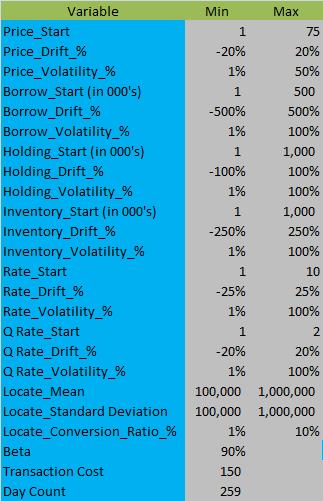}

\caption{Simulation Seed\label{fig:Simulation-Seed}}
\end{figure}

As noted earlier in section \ref{subsec:Valuation-Setup}, given the
number of random variables involved and the complexity of the system,
the computational infrastructure required to value an exclusive can
be tremendous. A typical exclusive portfolio can have anywhere from
a few hundred to upwards of a thousand different securities. It is
therefore simpler to use the historical time series and calculate
the valuation from the corresponding formula derived in section \ref{subsec:Historical-Valuations}.
To demonstrate numerical results we simulate the historical time series.
We pick a sample portfolio with one hundred different hypothetical
securities and we come up with the time series of all the variables
involved (Price, Quantity Borrowed, Exclusive Holding, Inventory Level,
Loan Rate, Alternate Loan Rate) by sampling from suitable log normal
distributions. It is worth noting that the mean and standard deviation
of each time series are themselves simulations from other appropriately
chosen uniform distributions (Figure \ref{fig:Simulation-Seed}).
The locate process can be modeled as a Poisson distribution with appropriately
chosen units. Though we consider the simpler alternative by letting
it be the absolute value of a normal distribution as justified in
section \ref{subsec:Valuation-Setup}. The mean and standard deviation
of the locate distribution for each security are chosen from another
appropriately chosen uniform distribution.

\begin{figure}[H]
\includegraphics[width=15cm]{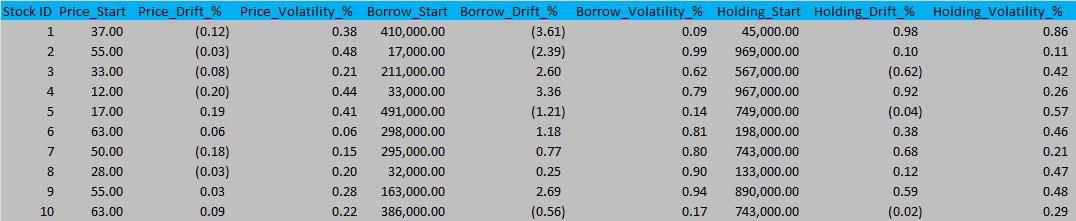}

\includegraphics[width=16.7cm]{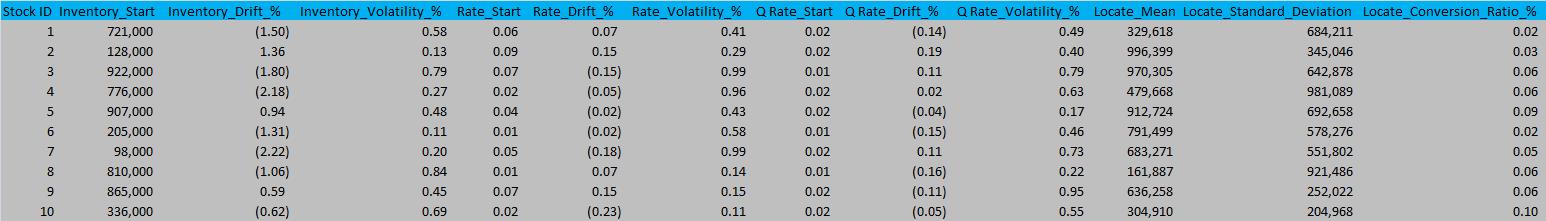}

\caption{Simulation Sample Distributions\label{fig:Simulation-Sample-Distributions}}
\end{figure}

\begin{figure}[H]
\includegraphics[width=7cm]{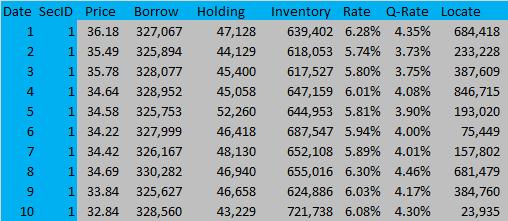}

\caption{Simulation Sample Time Series\label{fig:Simulation-Sample-Time}}
\end{figure}

The simulation seed is chosen so that the drift and volatility we
get for the variables (mean and standard deviation for the locate
process) are similar to what would be observed in practice. For example
in Figure \ref{fig:Simulation-Seed}, the price and rate volatility
are lower than the volatilities of the borrow and other quantities,
which tend to be much higher; the range of the drift for the quantities
is also higher as compared to the drift range of prices and rates.
 This ensures that we are keeping it as close to a realistic setting
as possible, without having access to the historical time series.
The volatility and drift of the variables for each security are shown
in Figure \ref{fig:Simulation-Sample-Distributions}. The length of
the simulated time series is one year or 252 trading days for each
security. A sample of the time series of the variables generated using
the simulated drift and volatility parameters is shown in Figure \ref{fig:Simulation-Sample-Time}.
The full time series used for the calculations is available upon request.


\subsection{\label{sec:Model-Testing-Results}Model Testing Results}

\begin{figure}[H]
\includegraphics[width=15cm]{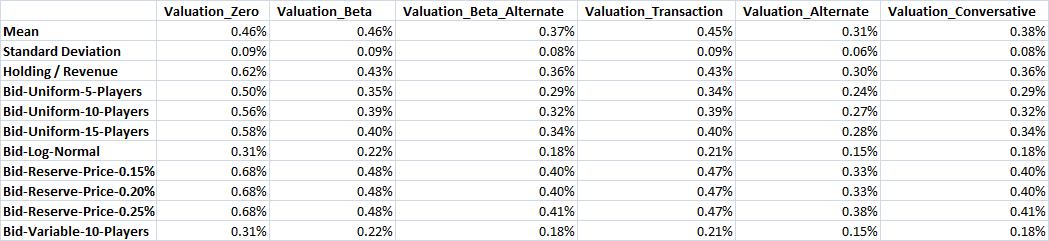}

\caption{Valuation Summary Statistics\label{fig:Valuation-Summary-Statistics}}

\end{figure}

To summarize the results of the testing we show the summary statistics
and the matrix of portfolio valuations under different valuation criteria
and auction settings in (Figure \ref{fig:Valuation-Summary-Statistics}).
The columns from one to six denote the following valuations $\left\{ \upsilon^{zero},\upsilon^{beta},\upsilon^{beta\;alternate},\upsilon^{transaction},\right.$
$\left.\upsilon^{alternate},\upsilon^{conservative}\right\} $ respectively.
The first and second rows are the mean and standard deviation of the
daily valuations. The third row indicates the valuation over the entire
time period under consideration. The fourth, fifth and sixth row indicate
the valuations when the valuations of the other bidders are distributed
uniformly and there are 5,10 or 15 other bidders respectively. The
seventh row indicates the log-normal assumption for the valuations
of the other bidders. The eight, ninth and tenth rows are when there
are reserve prices set by the auction seller of 15, 20 and 25 basis
points respectively with ten other bidders. The eleventh row indicates
the case when we have uncertainty about the number of bidders and
a total of ten bidders are distributed according to the discrete symmetric
distribution in (Figure \ref{fig:Variable-Bidders-Symmetric}). 

We see that the valuation ranges from 30 to 50 basis points under
different valuation schemes. The exclusive holding value varies between
1 billion to 2 billion over the time period under consideration. We
have not considered currency rates for simplicity, but a real portfolio
could hold securities traded in different currencies introducing foreign
exchange rate uncertainty into the mix. When we repeat the simulations
with different seed values, the results could vary outside this range,
but are not drastically different. The combined valuation based on
the result in theorem \ref{Th-1-When-each-of} is around 34 basis
points. This shows that the valuation with the lowest variance, which
is 30 basis points (Valuation\_Alternate: third row, fifth column
in Figure \ref{fig:Valuation-Summary-Statistics}) has a greater influence
on the combined valuation due to its higher weight in the aggregation.
This makes practical sense since a valuation with lesser variance
will provide more stable revenue streams. We have not provided the
auction bidding strategies based on this combined valuation, but these
can be easily calculated if someone wishes to proceed down that route.

\begin{figure}[H]
\includegraphics[width=11cm]{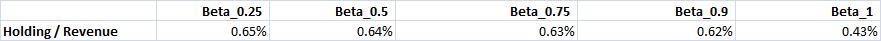}

\caption{Valuation Beta Comparative Statics\label{fig:Valuation-Beta-Comparative}}

\end{figure}

\begin{figure}[H]
\includegraphics[width=15cm]{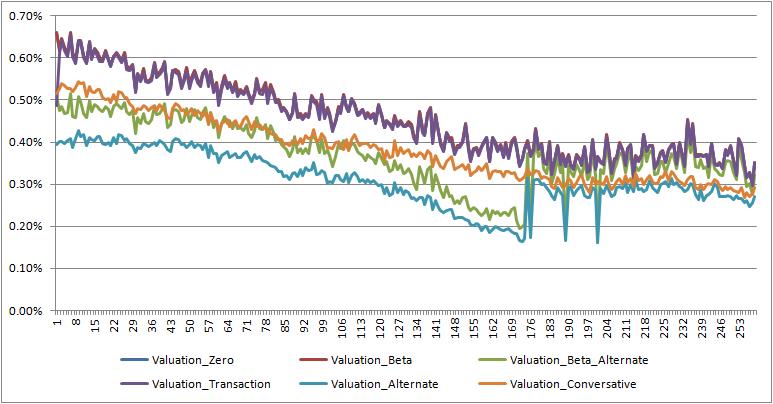}

\caption{Valuation Time Series\label{fig:Valuation-Time-Series}}
\end{figure}

We easily verify some results well known in the auction literature
(Krishna 2009): 1) As the number of bidders bidding uniformly increases,
the bid increases; 2) Setting a reserve price results in higher bids.
The bid with a discrete symmetric distribution as the number of bidders
goes higher is comparable to the crude theoretical approximation for
the log normal distribution, which does not depend on the number of
bidders (Corollary \ref{Corr-2-The-symmetric-equilibrium}). The Comparative
Statics of the valuation with changes in Beta are shown in Figures
\ref{fig:Valuation-Beta-Comparative}. As the subjective discount
factor $\beta$ decreases, the valuation increases since the effect
of the discounting is higher on the holding levels than on the revenue.
A time series graph of the different valuations are shown in Figure
\ref{fig:Valuation-Time-Series}. We have successfully utilized the
heuristic expression for the valuation of exclusives using real historical
data in making auction bids. As compared to the sample numbers shown
here with simulated data, when the historical observations are used
we get somewhat similar results.


\section{\label{sec:Improvements-to-the}Improvements to the Model}

Numerous improvements to the model are possible. (Cobb, Rumi \& Salmerón
2012; and Nie \& Chen 2007) derive approximate distributions for the
sum of log normal distribution which highlight that we can estimate
the log normal parameters from the time series of the valuations and
hence get the mean and variance of the valuations. A longer historical
time series will help to get better estimates for the volatility of
the valuation. This can be useful to decide the aggressiveness of
the bid. Another key extension can be to introduce jumps in the stochastic
processes. This is seen in stock prices to a certain extent and to
a greater extent in the borrow, holding and inventory processes. 

We have assumed that the stochastic processes governing the stocks
in the exclusive portfolio, the loan rates, the number of shares available
internally, as well as the number of shares available for loan from
other external borrowers are all independent. A more realistic assumption
of positive correlations between some of these processes would make
the results more realistic and appealing. Such a modification can
be incorporated into our framework but would require non trivial alternations
to the equations and numerical computations (Heston 1993; Oksendal
2013).

\begin{doublespace}
A key extension can be to introduce jumps in the stochastic processes
(Merton 1976; Kou 2002). This is seen in stock prices to a certain
extent (Yan 2011) and to a greater extent in the borrow, exclusive
holdings, availability and inventory processes. Another improvement
would be to use stochastic volatility for the stock prices (Hagan,
Kumar, Lesniewski \& Woodward 2002).
\end{doublespace}

The auction theory aspects combine standard results with new extensions
(Kashyap 2018) for the log-normal case, the interdependent case and
a combined realistic setting with uniform distributions. Instead of
the bidding strategies we have considered, we can come up with a parametric
model that will take the valuations as the inputs and the bid as output.
The parameters can depend on the size of the portfolio, the number
of securities, the number of special securities, the number of markets,
the extent of overlap with the internal inventory, and where available,
the percentile rankings of the historical bids for previous auctions,
which auction sellers do reveal sometimes. 

A significant amount of theoretical work could be pursued related
to the variance weighted combination of the valuations (section \ref{subsec:Variance-Weighted-Combined}).
In our example, we have considered valuations over the same time horizon
for the same portfolio. But extensions could look into combining valuations
or forecasts which are done over multiple horizons and even across
multiple objects with similar characteristics. 

A key open question is to decide which of the valuations to use for
the bidding strategy if we do not opt to combine them based on our
variance weighting (section \ref{subsec:Variance-Weighted-Combined}).
This aspect will require views on how the loan rates might evolve
and which securities in the exclusive pool will stay special or might
become special, and hence can be used to pick either a more aggressive
or a less aggressive valuation. In a subsequent paper (Kashyap 2016),
we will look at how we can systematically try and establish expectations
on loan rates and which securities might become harder to borrow and
hence have higher profit margins on the loans. The locate conversion
ratio can also be the result of profit maximization when the Knapsack
algorithm (Martello \& Toth 1987) is used to allocate the locates.

\section{\label{sec:Conclusion}Conclusion}

We have looked at methodologies to value securities portfolios from
a securities lending perspective. The weighting scheme we have used
to combine multiple valuations can be useful in any situation where
multiple forecasts are made and we need a methodology to combine them.
We have derived conditions under which our alternative weighting scheme
converges faster to the true valuation when compared to the minimum
variance weighting. We have then looked at various bidding strategies
that would be relevant to an exclusive auction. From a related paper
(Kashyap 2018), we have used closed form solutions where such a formulation
exists and in situations where approximations and numerical solutions
are required we have utilized those. The paper presents a theoretical
foundation supplemented with numerical results for a largely unexplored
financial business. The results from the simulation confirm the complexity
inherent in the system, but point out that the heuristics we have
used can be a practical tool for bidders and auction sellers to maximize
their profits. The models developed here could be potentially useful
for inventory estimation and for wholesale procurement of financial
instruments and also non-financial commodities. 

\section{\label{sec:Acknowledgments-and-End-notes}Acknowledgments and End-notes}
\begin{enumerate}
\item Dr. Yong Wang, Dr. Isabel Yan, Dr. Vikas Kakkar, Dr. Fred Kwan, Dr.
William Case, Dr. Srikant Marakani, Dr. Qiang Zhang, Dr. Costel Andonie,
Dr. Jeff Hong, Dr. Guangwu Liu, Dr. Andrew Chan, Dr. Humphrey Tung
and Dr. Xu Han at the City University of Hong Kong provided advice
and more importantly encouragement to explore and where possible apply
cross disciplinary techniques. The faculty members of SolBridge International
School of Business provided patient guidance and valuable suggestions
on how to further this paper.
\item Numerous seminar participants, particularly at a few meetings of the
econometric society and various finance organizations, provided helpful
suggestions. The views and opinions expressed in this article, along
with any mistakes, are mine alone and do not necessarily reflect the
official policy or position of either of my affiliations or any other
agency.
\item \label{enu:One}“An Introduction to Securities Lending” by Mark C
Faulkner, which may be downloaded at www.spitalfieldsadvisors.com,
was commissioned by the UK Securities Lending and Repo Committee,
the International Securities Lending Association, the London Stock
Exchange, the London Investment Banking Association, the British Bankers’
Association and the UK Association of Corporate Treasurers and was
welcomed by the National Association of Pension Funds and the Association
of British Insurers. It was first published in 2004.(\href{http://www.bankofengland.co.uk/markets/Documents/gilts/securitieslending.pdf}{Securities Lending One})
\item \label{enu:Two}“An Introduction to Securities Lending (Australia)”
is the Australian adaptation of the UK publication focused on the
UK markets, entitled “An Introduction to Securities Lending” by Mark
C Faulkner in (End-note \ref{enu:One}). This adaptation was commissioned
by the Australian Securities Lending Association Limited (ASLA). (\href{http://asla.com.au/info/LongSecLendingPaper.pdf}{Securities Lending Two})
\item \label{enu:Three}The value of available inventory as of June 22,
2015, stands at \$13.22 trillion, according to a new info-graphic
on the global securities finance market from DataLend. Of the available
inventory worldwide, \$1.72 trillion was out on loan. The value of
equity on loan was \$851 billion, while fixed income on loan stood
at \$876 billion. Some 41,673 unique securities were out loan, according
to the info-graphic, yielding an estimated gross revenue of \$19.2
million per day on average, which equates to \$2.26 billion for the
first half of 2015. The US is still the largest market with \$954
billion out on loan as of 22 June. Canada is the closest market in
size, with an estimated \$131 billion of securities out on loan. Despite
its size, the US commands a fee of 38 basis points (volume-weighted
average, year to date), whereas Hong Kong, which has \$28.8 billion
out on loan, yields fees of 210 basis points. (\href{http://www.securitieslendingtimes.com/securitieslendingnews/article.php?article_id=220006\#.VmahXL_O2iz}{Securites Lending Three})
\item \label{enu:Four}As the potential risks of securities lending are
discussed and debated by the Financial Stability Oversight Council
(FSOC), the U.S. Treasury’s Office of Financial Research (OFR), and
the Financial Stability Board (FSB), it is important to try to understand
both the overall size of the securities lending market and the share
of it attributable to different participants. Based on one estimate
from the FSOC the percentage is typically around these values (Retirement
and Pension, Mutual Funds, Endowments, Insurance: 50\%, 35\%, 8\%,
6\%). (\href{https://www.ici.org/viewpoints/ci.view_14_sec_lending_02.print}{Securites Lending Four}).
\item \label{enu:Despite-the-several}Despite the several advances in the
social sciences and in particular economic and financial theory, we
have yet to discover an objective measuring stick of value, a so called,
True Value Theory. While some would compare the search for such a
theory, to the medieval alchemist's obsession with turning everything
into gold, for our present purposes, the lack of such an objective
measure means that the difference in value as assessed by different
participants can effect a transfer of wealth. This forms the core
principle that governs all commerce that is not for immediate consumption
in general, and also applies specifically to all investment related
traffic which forms a great portion of the financial services industry.
Although, some of this is true for consumption assets; because the
consumption ability of individuals and organizations is limited and
their investment ability is not, the lack of an objective measure
of value affects investment assets in a greater way and hence investment
assets and related transactions form a much greater proportion of
the financial services industry. Consumption assets do not get bought
and sold, to an inordinate extent, due to fluctuating prices, whereas
investment assets will.
\item \textcolor{red}{\label{enu:=00005BComment-Only:-Lending}}The cyclical
nature of the transactions, which in some case can have its tentacles
spread far and wide, can result in catastrophic repercussions, especially
when huge sums of money move back and forth (Harrington 2009 is a
discussion of the financial crisis, systemic risk and the management
of collateral considering the role of American International Group,
AIG, as a case study; AIG ran into major problems with the securities
lending program of its life insurance subsidiaries when borrowers
requested the return of large amounts of collateral). No discussion
involving randomness is complete (Taleb 2005; 2010), especially one
involving randomness to the extent that we are tackling here, without
being highly attuned to spurious results mistakenly being treated
as correct and extreme situations causing devastating changes to the
expected outcomes.
\item \label{Triangular-Distribution}\href{https://en.wikipedia.org/wiki/Triangular_distribution}{Triangular distribution, Wikipedia Link}:
In probability theory and statistics, the triangular distribution
is a continuous probability distribution with lower limit $a$, upper
limit $b$ and mode $c$, where $a<b$ and $a\leq c\leq b$ (also
see: Evans, Hastings \& Peacock 2000).
\item \label{Irwin-Hall}\href{https://en.wikipedia.org/wiki/Irwin\%E2\%80\%93Hall_distribution}{Irwin–Hall distribution, Wikipedia Link}:
In probability and statistics, the Irwin--Hall distribution, named
after Joseph Oscar Irwin and Philip Hall, is a probability distribution
for a random variable defined as the sum of a number of independent
random variables, each having a uniform distribution. For this reason
it is also known as the uniform sum distribution (also see: Hall 1927;
Irwin 1927).
\item \label{Limit-Comparison-Test}\href{https://en.wikipedia.org/wiki/Limit_comparison_test}{Limit Comparison Test, WIkipedia Link}:
In mathematics, the limit comparison test (LCT) is a method of testing
for the convergence of an infinite series. Suppose that we have two
series $\Sigma_{n}a_{n}$ and $\Sigma_{n}b_{n}$ with ${\displaystyle a_{n}\geq0,b_{n}>0}$
for all $n$. Then if $\lim_{n\to\infty}\frac{a_{n}}{b_{n}}=c$ with
$0<c<\infty$ then either both series converge or both series diverge
(Swokowski 1979).
\item \label{Ratio-Test}\href{https://en.wikipedia.org/wiki/Ratio_test}{Ratio Test, Wikipedia Link}:
In mathematics, the ratio test is a test (or \textquotedbl criterion\textquotedbl )
for the convergence of a series $\sum_{{n=1}}^{\infty}a_{n}$, where
each term is a real or complex number and $a_{n}$ is nonzero when
$n$ is large. The usual form of the test makes use of the limit $L=\lim_{{n\to\infty}}\left|{\frac{a_{{n+1}}}{a_{n}}}\right|$.
The ratio test states that: if $L<1$ then the series converges absolutely
and $\lim_{{n\to\infty}}a_{n}\rightarrow0$; if $L>1$ then the series
is divergent; if $L=1$ or the limit fails to exist, then the test
is inconclusive, because there exist both convergent and divergent
series that satisfy this case (Bromwich 2005).
\item \label{Rate-Convergence}\href{https://en.wikipedia.org/wiki/Rate_of_convergence}{Rate of Convergence, Wikipedia Link}:
In numerical analysis, the speed at which a convergent sequence approaches
its limit is called the rate of convergence. Suppose that the sequence
${\displaystyle (x_{k})_{k}}$ converges to the number $L$. This
sequence is said to converge linearly to $L$, if there exists a number
${\displaystyle \mu\in(0,1)}$ such that ${\displaystyle \lim_{k\to\infty}{\frac{|x_{k+1}-L|}{|x_{k}-L|}}=\mu.}$
The number $\mu$ is called the rate of convergence. The smaller the
number $\mu$, faster the convergence (Schatzman 2002).
\item \label{Product of Sum and their Reciprocals}\href{http://www.stumblingrobot.com/2015/07/11/prove-the-product-of-sums-of-n-numbers-and-their-reciprocals-is-greater-than-n2/}{Product of Sums of Numbers and their Reciprocals}:
We apply the Cauchy-Schwarz inequality to the numbers $\sqrt{x_{1}},\ldots,\sqrt{x_{n}}$
and $\sqrt{y_{1}},\ldots,\sqrt{y_{n}}$. Here, $x_{1},\ldots,x_{n}\in\mathbb{R}_{>0}$
and $y_{k}=\frac{1}{x_{k}}$ for $k=1,\ldots,n$. Since, $y_{k}=\frac{1}{x_{k}}\;\implies\;\sqrt{y_{k}}=\frac{1}{\sqrt{x_{k}}}\;\implies\;\sqrt{x_{k}}\sqrt{y_{k}}=1$.
We then get $\left(\sum_{k=1}^{n}x_{k}\right)\left(\sum_{k=1}^{n}y_{k}\right)\geq n^{2}.$
\end{enumerate}

\section{References }
\begin{enumerate}
\item Aitken, M. J., Frino, A., McCorry, M. S., \& Swan, P. L. (1998). Short
sales are almost instantaneously bad news: Evidence from the Australian
Stock Exchange. The Journal of Finance, 53(6), 2205-2223.
\item Allen, F., Morris, S., \& Postlewaite, A. (1993). Finite bubbles with
short sale constraints and asymmetric information. Journal of Economic
Theory, 61(2), 206-229.
\item Baklanova, V., Copeland, A., \& McCaughrin, R. (2015). Reference Guide
to US Repo and Securities Lending Markets (No. 740).
\item Bates, J. M., \& Granger, C. W. (1969). The combination of forecasts.
Journal of the Operational Research Society, 20(4), 451-468.
\item Boehmer, E., Jones, C. M., \& Zhang, X. (2008). Which shorts are informed?.
The Journal of Finance, 63(2), 491-527.
\item Bris, A., Goetzmann, W. N., \& Zhu, N. (2007). Efficiency and the
bear: Short sales and markets around the world. The Journal of Finance,
62(3), 1029-1079.
\item Bromwich, T. J. I. A. (2005). An Introduction to the Theory of Infinite
Series (Vol. 335). American Mathematical Soc..
\item Campbell, J. Y., Lo, A. W., MacKinlay, A. C., \& Whitelaw, R. F. (1998).
The econometrics of financial markets. Macroeconomic Dynamics, 2(04),
559-562.
\item Chan, F., \& Pauwels, L. L. (2018). Some theoretical results on forecast
combinations. International Journal of Forecasting, 34(1), 64-74.
\item Cheng, T. T. (1949). The normal approximation to the Poisson distribution
and a proof of a conjecture of Ramanujan. Bulletin of the American
Mathematical Society, 55(4), 396-401.
\item Chiani, M., Dardari, D., \& Simon, M. K. (2003). New exponential bounds
and approximations for the computation of error probability in fading
channels. Wireless Communications, IEEE Transactions on, 2(4), 840-845.
\item Clemen, R. T. (1989). Combining forecasts: A review and annotated
bibliography. International journal of forecasting, 5(4), 559-583.
\item Cobb, B. R., Rumi, R., \& Salmerón, A. (2012). Approximating the Distribution
of a Sum of Log-normal Random Variables. Statistics and Computing,
16(3), 293-308.
\item Cochrane, J. H. (2009). Asset Pricing:(Revised Edition). Princeton
university press.
\item Cohen, L., Diether, K. B., \& Malloy, C. J. (2007). Supply and demand
shifts in the shorting market. The Journal of Finance, 62(5), 2061-2096.
\item Conflitti, C., De Mol, C., \& Giannone, D. (2015). Optimal combination
of survey forecasts. International Journal of Forecasting, 31(4),
1096-1103.
\item D’avolio, G. (2002). The market for borrowing stock. Journal of financial
economics, 66(2), 271-306.
\item Desai, H., Ramesh, K., Thiagarajan, S. R., \& Balachandran, B. V.
(2002). An investigation of the informational role of short interest
in the Nasdaq market. The Journal of Finance, 57(5), 2263-2287.
\item Diamond, D. W., \& Verrecchia, R. E. (1987). Constraints on short-selling
and asset price adjustment to private information. Journal of Financial
Economics, 18(2), 277-311.
\item Duffie, D., Garleanu, N., \& Pedersen, L. H. (2002). Securities lending,
shorting, and pricing. Journal of Financial Economics, 66(2), 307-339.
\item Dyer, D., Kagel, J. H., \& Levin, D. (1989). Resolving uncertainty
about the number of bidders in independent private-value auctions:
an experimental analysis. The RAND Journal of Economics, 268-279.
\item Erdos, P., \& Hunt, G. A. (1953). Changes of sign of sums of random
variables. Pacific J. Math, 3, 673-687.
\item Etemadi, N. (2006). Convergence of weighted averages of random variables
revisited. Proceedings of the American Mathematical Society, 134(9),
2739-2744.
\item Evans, M., Hastings, N., \& Peacock, B. (2000). Triangular distribution.
Statistical distributions, 3, 187-188.
\item Granger, C. W., \& Ramanathan, R. (1984). Improved methods of combining
forecasts. Journal of forecasting, 3(2), 197-204.
\item Granger, C. W. (1989). Invited review combining forecasts---twenty
years later. Journal of Forecasting, 8(3), 167-173.
\item Gujarati, D. N. (1995). Basic econometrics, 3rd. International Edition.
\begin{doublespace}
\item Hagan, P. S., Kumar, D., Lesniewski, A. S. and Woodward, D. E. (2002)
Managing smile risk, WILMOTT Magazine, September, pp. 84--108.
\end{doublespace}
\item Hamilton, J. D. (1994). Time series analysis (Vol. 2). Princeton university
press.
\item Harrington, S. E. (2009). The financial crisis, systemic risk, and
the future of insurance regulation. Journal of Risk and Insurance,
76(4), 785-819.
\item Harrison, J. M., \& Kreps, D. M. (1978). Speculative investor behavior
in a stock market with heterogeneous expectations. The Quarterly Journal
of Economics, 323-336.
\item Harstad, R. M., Kagel, J. H., \& Levin, D. (1990). Equilibrium bid
functions for auctions with an uncertain number of bidders. Economics
Letters, 33(1), 35-40.
\item Heston, S. L. (1993). A closed-form solution for options with stochastic
volatility with applications to bond and currency options. The review
of financial studies, 6(2), 327-343.
\item Hong, H., \& Stein, J. C. (2003). Differences of opinion, short-sales
constraints, and market crashes. Review of financial studies, 16(2),
487-525.
\item Hsiao, C., \& Wan, S. K. (2014). Is there an optimal forecast combination?.
Journal of Econometrics, 178, 294-309.
\item Hull, J. C. (2010). Options, Futures, and Other Derivatives, 7/e (With
CD). Pearson Education India.
\item Jarrow, R. (1980). Heterogeneous expectations, restrictions on short
sales, and equilibrium asset prices. The Journal of Finance, 35(5),
1105-1113.
\item Jones, C. M., \& Lamont, O. A. (2002). Short-sale constraints and
stock returns. Journal of Financial Economics, 66(2), 207-239.
\begin{doublespace}
\item Kashyap, R. (2016). Securities Lending Strategies, TBR and TBR (Theoretical
Borrow Rate and Thoughts Beyond Rates). Social Science Research Network
(SSRN), Working Paper.
\item Kashyap, R. (2017). Notes on Uncertainty, Unintended Consequences
and Everything Else. Social Science Research Network (SSRN), Working
Paper.
\end{doublespace}
\item Kashyap, R. (2018). Auction theory adaptations for real life applications.
Research in Economics, 72(4), 452-481.
\item Klemperer, P. (2004). Auctions: theory and practice.
\item Kolasinski, A. C., Reed, A. V., \& Ringgenberg, M. C. (2013). A multiple
lender approach to understanding supply and search in the equity lending
market. The Journal of Finance, 68(2), 559-595.
\begin{doublespace}
\item Kou, S. G. (2002). A jump-diffusion model for option pricing. Management
science, 48(8), 1086-1101.
\end{doublespace}
\item Krishna, V. (2009). Auction theory. Academic press.
\item Laffont, J. J., Ossard, H., \& Vuong, Q. (1995). Econometrics of first-price
auctions. Econometrica: Journal of the Econometric Society, 953-980.
\item Lai, T. L., \& Xing, H. (2008). Statistical models and methods for
financial markets. New York: Springer.
\item Lebrun, B. (1999). First price auctions in the asymmetric N bidder
case. International Economic Review, 40(1), 125-142.
\item Levin, D., \& Ozdenoren, E. (2004). Auctions with uncertain numbers
of bidders. Journal of Economic Theory, 118(2), 229-251.
\item Loeve, M. (1977). Elementary probability theory. In Probability Theory
I (pp. 1-52). Springer, New York, NY.
\item Martello, S., \& Toth, P. (1987). Algorithms for knapsack problems.
Surveys in combinatorial optimization, 31, 213-258.
\item Menezes, F. M., \& Monteiro, P. K. (2005). An introduction to auction
theory (pp. 1-178). Oxford University Press.
\begin{doublespace}
\item Merton, R. C. (1976). Option pricing when underlying stock returns
are discontinuous. Journal of financial economics, 3(1-2), 125-144.
\end{doublespace}
\item Milgrom, P. R. (2004). Putting auction theory to work. Cambridge University
Press.
\item Milgrom, P. R., \& Weber, R. J. (1982). A theory of auctions and competitive
bidding. Econometrica: Journal of the Econometric Society, 1089-1122.
\begin{doublespace}
\item Miranda, M. J., \& Fackler, P. L. (2002). Applied Computational Economics
and Finance.
\end{doublespace}
\item Mitchell, M., Pulvino, T., \& Stafford, E. (2002). Limited arbitrage
in equity markets. The Journal of Finance, 57(2), 551-584.
\item Morris, S. (1996). Speculative investor behavior and learning. The
Quarterly Journal of Economics, 1111-1133.
\item Nie, H., \& Chen, S. (2007). Lognormal sum approximation with type
IV Pearson distribution. IEEE Communications Letters, 11(10), 790-792.
\item Norstad, John. \textquotedbl The normal and lognormal distributions.\textquotedbl{}
(1999).
\item Ofek, E., \& Richardson, M. (2003). Dotcom mania: The rise and fall
of internet stock prices. The Journal of Finance, 58(3), 1113-1137.
\item Ofek, E., Richardson, M., \& Whitelaw, R. F. (2004). Limited arbitrage
and short sales restrictions: Evidence from the options markets. Journal
of Financial Economics, 74(2), 305-342.
\item Oksendal, B. (2013). Stochastic differential equations: an introduction
with applications. Springer Science \& Business Media.
\item Ortega-Reichert, A. (1967). Models for competitive bidding under uncertainty.
Stanford University.
\item Reed, A. (2007). Costly short-selling and stock price adjustment to
earnings announcements. University of North Carolina, unpublished
manuscript.
\item Schatzman, M. (2002). Numerical analysis: a mathematical introduction.
Oxford University Press.
\item Scheinkman, J. A., \& Xiong, W. (2003). Overconfidence and speculative
bubbles. Journal of political Economy, 111(6), 1183-1220.
\item Swokowski, E. W. (1979). Calculus with analytic geometry. Taylor \&
Francis.
\item Smith, J., \& Wallis, K. F. (2009). A simple explanation of the forecast
combination puzzle. Oxford Bulletin of Economics and Statistics, 71(3),
331-355.
\item Sweeney, Joan, and Richard James Sweeney. \textquotedbl Monetary
theory and the great Capitol Hill Baby Sitting Co-op crisis: comment.\textquotedbl{}
Journal of Money, Credit and Banking 9.1 (1977): 86-89.
\item Taleb, Nassim. Fooled by randomness: The hidden role of chance in
life and in the markets. Random House Trade Paperbacks, 2005.
\item Taleb, Nassim Nicholas. The Black Swan: The Impact of the Highly Improbable
Fragility. Random House Digital, Inc., 2010.
\item Timmermann, A. (2006). Forecast combinations. Handbook of economic
forecasting, 1, 135-196.
\begin{doublespace}
\item Yan, S. (2011). Jump risk, stock returns, and slope of implied volatility
smile. Journal of Financial Economics, 99(1), 216-233.
\end{doublespace}
\end{enumerate}

\section{\label{sec:Securities-Lending-Background}Appendix: Securities Lending
Background}

Securities Lending began as an informal practice among brokers who
had insufficient share certificates to settle their sold bargains,
commonly because their selling clients had misplaced their certificates
or just not provided them to the broker by the settlement date of
the transaction. Once the broker had received the certificates, they
would be passed on to the lending broker. This arrangement was not
subject to any formal agreements and there was no exchange of collateral.
Today, securities lending is a significant market practice whereby
securities are temporarily transferred by one party, (the lender)
to another (the borrower). The borrower is obliged to return the securities
to the lender, either on demand or at the end of any agreed term.
For the period of the loan, the lender is secured by acceptable assets
or cash of equal or greater value than the lent securities, delivered
by the borrower to the lender, as collateral. With such simple beginnings,
today the business generates hundreds of millions of dollars in revenue
and involves the movement of trillions of dollars’ worth of financial
instruments. The Over-The-Counter (OTC) nature of the business means
that is hard to come up with actual numbers in terms of size and profitability,
but we provide some estimates in section \ref{subsec:Exclusive-Auctions-Wallet}. 

Below we chronicle various circumstances that lead to the demand for
securities loans.
\begin{enumerate}
\item Market making and proprietary trading
\begin{itemize}
\item The most common reason to borrow securities is to cover a short position
using the borrowed securities to settle an outright sale. But this
is rarely a simple speculative bet that the value of a security will
fall, so that the borrower can buy it more cheaply at the maturity
of the loan. More commonly, the short position is part of a larger
trading strategy, typically designed to profit from perceived pricing
discrepancies between related securities. Some examples are:
\begin{itemize}
\item Convertible bond arbitrage: buying a convertible bond and simultaneously
selling the underlying equity short.
\item ‘Pairs’ trading: seeking to identify two companies, with similar characteristics,
whose equity securities are currently trading at a price relationship
that is out of line with the historical trading range. The apparently
undervalued security is bought, while the apparently overvalued security
is sold short.
\item Merger arbitrage: for example, selling short the equities of a company
making a takeover bid against a long position in those of the potential
acquisition company.
\item Index arbitrage: selling short the constituent securities of an equity
price index against a long position in the corresponding index future.
\item Other market making and proprietary trading related activities that
require borrowing securities include equity / derivative arbitrage,
and equity option hedging.
\end{itemize}
\end{itemize}
\item Borrowing for Failed Trades
\begin{itemize}
\item A failed trade may be defined as one where delivery cannot be completed
because of insufficient securities available. This is not deliberate
policy, but is caused by any number of general administrative problems.
Borrowings to cover fails are mostly small and short in duration (one
to five days). The borrower keeps the loan open only until he can
complete delivery of the underlying trade. An example of this type
of transaction occurs when a broker’s client sells stock, but fails
to deliver the securities to his broker. The broker borrows the stock,
settles the trade and places the resultant settlement funds on deposit.
He thereby earns interest on this cash and avoids fail fines. He then
unwinds the loan once the client has delivered his securities.
\end{itemize}
\item Borrowing for Margin Requirements.
\begin{itemize}
\item To meet margin requirements, for example at the exchange traded options
market, securities can be borrowed cheaply and lodged as margin, rather
than depositing cash.
\end{itemize}
\item Temporary Transfer of Ownership
\begin{itemize}
\item Another large class of transactions not involving a short is motivated
by lending to transfer ownership temporarily to the advantage of both
lender and borrower. For example, where a lender would be subject
to withholding tax on dividends or interest but some potential borrowers
are not. Subject to the possible application of any relevant specific
or general anti-avoidance tax provisions or principles, the borrower
receives the dividend free of tax and shares some of the benefit with
the lender in the form of a larger fee or larger manufactured dividend.
\end{itemize}
\end{enumerate}
Security loans drawn down by market makers and traders on equity instruments
are typified as being large in volume and long in duration. For lenders,
these loans represent the greatest opportunity to maximize profit.
This is also the reason for referring to these business units as stock
loan desks, even though they lend fixed income securities, handle
repurchase agreements, manage collateral and other securities borrowing
related activities. 

The supply of securities into the lending market comes mainly from
the portfolios of beneficial owners such as pensions, insurance companies
and other such funds. Majority of the funds or asset owners work through
agents or intermediary brokers. Intermediaries act between lenders
and borrowers. For their services, the intermediary takes a spread.
Many institutions find it convenient to lend stock to one or two intermediaries
who then lend on to many more counter-parties. This saves administrative
overheads and limits credit risks. The spread is the result of a bargaining
process between intermediary brokers and beneficial owners on one
side and between intermediary brokers and end borrowers on the other
side.

\section{\label{sec:Appendix:-Dictionary-of}Appendix: Dictionary of Notation
and Terminology for the Exclusive Valuation}
\begin{itemize}
\item $B_{it}$, the Loan Book carried by the desk, in shares, at a particular
time, $t$, for security, $i$. This is the existing amount borrowed
by external lenders and hence can also be termed the borrow book.
\item $L_{it}$, the Locate Requests received, in shares, at a particular
time, $t$, for security, $i$.
\item $\delta_{it}\in\left[0,1\right],$ the conversion rate of locates
into borrows, at a particular time, $t$, for security, $i$. We can
simplify this to be the same per security.
\item $\delta_{i},$ the conversion rate of locates into borrows for security,
$i$. We can simplify this further to be a constant across time and
securities, $\delta$.
\item $\delta_{i}$$L_{it}$, then indicates the excess demand that the
desk receives, in shares, at a particular time, $t$, for security,
$i$.
\item $I_{it}$, the Internal Inventory the intermediary holds, in shares,
at a particular time, $t$, for security, $i$.
\item $O_{it}$, the additional supply that can be sourced from beneficial
owners other than the exclusive, in shares, at a particular time,
$t$, for security, $i$.
\item $A_{it}$, the Amount taken out from the Exclusive pool, in shares,
at a particular time, $t$, for security, $i$.
\item $H_{it}$, the Holdings available in the Exclusive pool, in shares,
at a particular time, $t$, for security, $i$.
\item $R_{it}$, the Rate on the loan charged by the intermediary, at a
particular time, $t,$ until the next time period, $t+1$, for security,
$i.$
\item $Q_{it}$, an alternate rate to $R_{it}$, at a particular time, t,
until the next time period, $t+1$, for security, $i$. This could
be the rate at which supply from other beneficial owners is sourced
or could be theoretical rate when no rate from other beneficial owners
is available. $Q_{it}\leq R_{it}$.
\item $S_{it}$, the Security Price at a particular time, $t,$ until the
next time period, $t+1$, for security, $i.$
\item $\beta=\frac{1}{\left(1+s\right)},$ is the discount factor, $s$
is the risk free rate of interest. Further complications can be introduced
by incorporating continuous time extensions to the short rate process. 
\item $\upsilon,$ the Valuation of the exclusive, for the duration extending
from $t=0\;to\;t=T.$
\item The total duration for which the exclusive will be contracted, $T$.
\item $P$, the profits from the exclusive for the intermediary over the
entire duration $T$.
\item $T_{s}$ and $T_{e}$ are the start and end times of the historical
time series.
\item $n,$ the number of securities available in the Exclusive pool, $i\in\left\{ 1,\,...\,,n\right\} $.
\item $c,$ the transaction cost each time shares are taken or put back
into the exclusive.
\item $N,$ the number of trading intervals.
\item The length of each trading interval, $\tau=T/N$. We assume the time
intervals are of the same duration, but this can be relaxed quite
easily. \\
In continuous time, this becomes, $N\rightarrow\infty,\tau\rightarrow0$.
\item The time then becomes divided into discrete intervals, $t_{k}=k\tau,\;k=0,...,N$.
We simplify this and write it as $t=0\;to\;t=T$ with unit increments.
\item It is common practice to consider daily increments in time for one
year. The fees paid generally also applies on weekends and holidays,
though there would be no change in any of the variables on these days.
Some firms use 252 trading days to annualize daily loan rates and
other fee terms. 
\item $\left\{ \upsilon^{zero},\upsilon^{beta},\upsilon^{beta\;alternate},\upsilon^{transaction},\right.$
$\left.\upsilon^{conservative},\upsilon^{alternate},\upsilon^{historical}\right\} $,
is the set of valuations.
\end{itemize}

\section{\label{sec:Appendix-A:-Valuation}Appendix: Valuation Proofs}

\subsection{\label{subsec:Proof-of-Proposition-Zero-Profits}Proof of Proposition
\ref{1-The-zero-profits}}
\begin{proof}
First, we simplify the constraints by reasoning as follows. If there
is other external supply, $O_{it}$ , being used, then we have 
\begin{eqnarray*}
I_{it}+H_{it} & \leq & B_{it}+\delta_{i}L_{it}\\
\Rightarrow I_{it}+A_{it} & \leq & B_{it}+\delta_{i}L_{it}\\
\Rightarrow A_{it} & \leq & B_{it}+\delta_{i}L_{it}-I_{it}\\
\Rightarrow\text{if}\;B_{it}+\delta_{i}L_{it} & \leq & I_{it}\;\text{then}\;A_{it}=0
\end{eqnarray*}
The maximum possible value of $A_{it}$ is then given by
\[
A_{it}=\min\left[H_{it},\max\left(B_{it}+\delta_{i}L_{it}-I_{it},0\right)\right]
\]
The criteria for zero profits, gives us an expression for the maximum
possible value of the exclusive.
\begin{eqnarray*}
P & = & \underset{A_{it}}{\max}\;E_{0}\left\{ \sum_{t=0}^{T}\beta^{t}\sum_{i=1}^{n}A_{it}S_{it}R_{it}-\sum_{t=0}^{T}\upsilon\beta^{t}\left(\sum_{i=1}^{n}H_{it}S_{it}\right)\right\} \\
 & = & \;E_{0}\left\{ \sum_{t=0}^{T}\sum_{i=1}^{n}\beta^{t}\min\left[H_{it},\max\left(B_{it}+\delta_{i}L_{it}-I_{it},0\right)\right]S_{it}R_{it}-\sum_{t=0}^{T}\upsilon\beta^{t}\left(\sum_{i=1}^{n}H_{it}S_{it}\right)\right\} 
\end{eqnarray*}
\[
\Rightarrow\upsilon^{actual}=\upsilon\leq\upsilon^{zero}=E_{0}\left\{ \frac{\sum_{t=0}^{T}\sum_{i=1}^{n}\beta^{t}\min\left[H_{it},\max\left(B_{it}+\delta_{i}L_{it}-I_{it},0\right)\right]S_{it}R_{it}}{\sum_{t=0}^{T}\beta^{t}\left(\sum_{i=1}^{n}H_{it}S_{it}\right)}\right\} 
\]
\end{proof}

\subsection{\label{subsec:Proof-of-Proposition-2-Transaction-Costs}Proof of
Proposition \ref{2-The-valuation-expression}}
\begin{proof}
Let the following variables, $\left\{ Take_{t},Give_{t}\right\} $
represented by the corresponding functions below, denote the criteria
that captures when there would be a need to take from (when the total
demand including existing borrows and a portion of the locate requests
is more than the internal inventory) or give back to the exclusive. 

\begin{eqnarray*}
High\;State\equiv Take_{t} & \equiv & \frac{max\left(B_{it}+\delta_{i}L_{it}-I_{it},0\right)}{\left(B_{it}+\delta_{i}L_{it}-I_{it}\right)}=\begin{cases}
1\;if\; & B_{it}+\delta_{i}L_{it}>I_{it}\\
0 & Otherwise
\end{cases}\\
Low\;State\equiv Give_{t} & \equiv & \frac{max\left(I_{it}-B_{it}-\delta_{i}L_{it},0\right)}{\left(I_{it}-B_{it}-\delta_{i}L_{it}\right)}=\begin{cases}
1\;if\; & B_{it}+\delta_{i}L_{it}<I_{it}\\
0 & Otherwise
\end{cases}
\end{eqnarray*}
It is worth noting that $Take_{t}$ and $Give_{t}$ are mutually exclusive.
Only one of them can be one in a given time period. We consider the
following four scenarios that can happen, back to back, or in successive
time periods. 
\[
\left[\left\{ Take_{t-1},Take_{t}\right\} \left\{ Take_{t-1},Give_{t}\right\} \left\{ Give_{t-1},Give_{t}\right\} \left\{ Give_{t-1},Take_{t}\right\} \right]
\]
Of the above scenarios, the following indicates the transaction cost
incurred correspondingly. There is a cost, when a state change occurs
either from Take to Give or from Give to Take.
\[
\left[\left\{ 0\right\} \left\{ c\right\} \left\{ 0\right\} \left\{ c\right\} \right]
\]
The above is equivalent to 
\[
\left[\left\{ Take_{t},Take_{t-1}\right\} \left\{ Take_{t},Give_{t-1}\right\} \left\{ Give_{t},Give_{t-1}\right\} \left\{ Give_{t},Take_{t-1}\right\} \right]\equiv\left[\left\{ 0\right\} \left\{ c\right\} \left\{ 0\right\} \left\{ c\right\} \right]
\]
The table below (Figure \ref{fig:Transacion-Cost-Table}) summarizes
the transaction costs incurred, based on the difference between variables
across successive time periods, when one of the four combinations
occurs. As an example, $\left\{ Give_{t},Take_{t-1}\right\} $ means
that in time period $t-1$, the system is in the $High\;State$, or
$Take_{t-1}=1$ and in time period $t$ it is in the $Low\;State$,
or $Give_{t}=1$. Hence, when this combination occurs, we have, $c\left(Give_{t}-Give_{t-1}\right)=c$
and $c\left(Take_{t}-Take_{t-1}\right)=-c$.

{\footnotesize{}}
\begin{figure}[H]
\begin{tabular}{|l|c|c|c|c|}
\hline 
 & $\left\{ Take_{t},Take_{t-1}\right\} $ & $\left\{ Take_{t},Give_{t-1}\right\} $ & $\left\{ Give_{t},Give_{t-1}\right\} $ & $\left\{ Give_{t},Take_{t-1}\right\} $\tabularnewline
\hline 
\hline 
$c\left(Take_{t}-Take_{t-1}\right)$ & $0$ & $c$ & $0$ & $-c$\tabularnewline
\hline 
$c\left(Take_{t}-Give_{t-1}\right)$ & $c$ & $0$ & $-c$ & $0$\tabularnewline
\hline 
$c\left(Give_{t}-Give_{t-1}\right)$ & $0$ & $-c$ & $0$ & $c$\tabularnewline
\hline 
$c\left(Give_{t}-Take_{t-1}\right)$ & $-c$ & $0$ & $c$ & $0$\tabularnewline
\hline 
\end{tabular}

\caption{{\footnotesize{}\label{fig:Transacion-Cost-Table}}Transaction Cost
Table}
\end{figure}
From this we get the expression for the transaction costs incurred,
keeping in mind that in the first time period, High State or Take
criterion would always incur a cost.
\begin{eqnarray*}
Transaction\;Costs\;\equiv TC & = & E_{0}\left\{ \sum_{i=1}^{n}c\left\{ Take_{i0}\right\} \right.\\
 & + & \left.\sum_{t=1}^{T}\sum_{i=1}^{n}\frac{c}{2}\left[\left|\left\{ Take_{it}-Take_{it-1}\right\} -\left\{ Give_{it}-Give_{it-1}\right\} \right|\right]\right\} \\
\Rightarrow TC & = & E_{0}\left\{ \sum_{i=1}^{n}c\left\{ \frac{max\left(B_{i0}+\delta_{i}L_{i0}-I_{i0},0\right)}{\left(B_{i0}+\delta_{i}L_{i0}-I_{i0}\right)}\right\} \right.\\
 & + & \sum_{t=1}^{T}\sum_{i=1}^{n}\frac{c}{2}\left[\left|\left\{ \frac{max\left(B_{it}+\delta_{i}L_{it}-I_{it},0\right)}{\left(B_{it}+\delta_{i}L_{it}-I_{it}\right)}\right.\right.\right.\\
 & - & \left.\frac{max\left(B_{it-1}+\delta_{i}L_{it-1}-I_{it-1},0\right)}{\left(B_{it-1}+\delta_{i}L_{it-1}-I_{it-1}\right)}\right\} \\
 & - & \left\{ \frac{max\left(I_{it}-B_{it}-\delta_{i}L_{it},0\right)}{\left(I_{it}-B_{it}-\delta_{i}L_{it}\right)}\right.\\
 & - & \left.\left.\left.\left.\frac{max\left(I_{it-1}-B_{it-1}-\delta_{i}L_{it-1},0\right)}{\left(I_{it-1}-B_{it-1}-\delta_{i}L_{it-1}\right)}\right\} \right|\right]\right\} 
\end{eqnarray*}
\begin{eqnarray*}
\Rightarrow\upsilon^{transaction} & = & E_{0}\left\{ \frac{\left(\sum_{t=0}^{T}\sum_{i=1}^{n}min\left[H_{it},max\left(B_{it}+\delta_{i}L_{it}-I_{it},0\right)\right]S_{it}R_{it}\right)-\left(TC\right)}{\left(\sum_{t=0}^{T}\sum_{i=1}^{n}H_{it}S_{it}\right)}\right\} 
\end{eqnarray*}
\end{proof}

\subsection{\label{subsec:Proof-of-Theorem-Variance-Weighted}Proof of Theorem
\ref{Th-1-When-each-of}}

\subsubsection{\label{subsec:Asymptotic-Expected-Value}Asymptotic Expected Value
of the Alternative Weighting Scheme}
\begin{proof}
We proceed by taking the expectation of the alternative weighting
scheme with the limit, $k\rightarrow\infty$, as follows, 

\[
E\left[\underset{k\rightarrow\infty}{\lim}\frac{1}{\left(k\right)}\frac{\underset{i=1}{\overset{k}{\sum}}\;\underset{j\neq i}{\overset{k}{\sum}}\;\sigma_{j}^{2}\upsilon_{i}}{\underset{i=1}{\overset{k}{\sum}}\sigma_{i}^{2}}\right]=E\left[\underset{k\rightarrow\infty}{\lim}\frac{1}{\left(k\right)}\frac{\underset{i=1}{\overset{k}{\sum}}\;\left(\underset{j=1}{\overset{k}{\sum}}\sigma_{j}^{2}\upsilon_{i}-\sigma_{i}^{2}\upsilon_{i}\right)}{\underset{i=1}{\overset{k}{\sum}}\sigma_{i}^{2}}\right]
\]
\[
=E\left[\underset{k\rightarrow\infty}{\lim}\frac{1}{\left(k\right)}\frac{\underset{i=1}{\overset{k}{\sum}}\;\underset{j=1}{\overset{k}{\sum}}\sigma_{j}^{2}\upsilon_{i}-\underset{i=1}{\overset{k}{\sum}}\;\sigma_{i}^{2}\upsilon_{i}}{\underset{i=1}{\overset{k}{\sum}}\sigma_{i}^{2}}\right]
\]
\[
=E\left[\underset{k\rightarrow\infty}{\lim}\frac{1}{\left(k\right)}\left\{ \underset{i=1}{\overset{k}{\sum}}\left(\upsilon_{i}\right)-\frac{\underset{i=1}{\overset{k}{\sum}}\;\sigma_{i}^{2}\upsilon_{i}}{\underset{i=1}{\overset{k}{\sum}}\sigma_{i}^{2}}\right\} \right]
\]
We consider two cases depending on whether the valuations have the
same finite expected value or not. We note that the following conditions
are common to both the cases below.
\[
\left\{ \quad\quad\vphantom{\left[\underset{k\rightarrow\infty}{\lim}\frac{\max\left(\sigma_{i}^{2}\right)}{\underset{i=1}{\overset{k}{\sum}}\;\sigma_{i}^{2}}\rightarrow0,\right]}\underset{k\rightarrow\infty}{\lim}\frac{1}{\left(k\right)}\frac{\underset{i=1}{\overset{k}{\sum}}\;\sigma_{i}^{2}\upsilon_{i}}{\underset{i=1}{\overset{k}{\sum}}\sigma_{i}^{2}}=0,\text{ Since each of the variances and valuations are finite, }\right.
\]
\begin{eqnarray*}
\text{ and the following combination of valuations and variances is uniformly bounded, that is, }
\end{eqnarray*}
\[
\underset{k\rightarrow\infty}{\lim}\frac{\left(\sum_{i=1}^{k}\sigma_{i}^{2}v_{i}\right)}{(\sum_{i=1}^{k}\sigma_{i}^{2})}\leq M_{1},\text{ any real number }
\]
\begin{eqnarray*}
\text{ and the covariance between the valuations is zero, }cov\left(v_{i},v_{i}\right)=0;\frac{(\sum_{i=1}^{k}\sigma_{i}^{2})}{k^{2}}\rightarrow0\text{ as }k\rightarrow\infty
\end{eqnarray*}
\begin{eqnarray*}
\left.\text{ and no single one dominates the sum, expressed as, }\left[\underset{k\rightarrow\infty}{\lim}\frac{\max\left(\sigma_{i}^{2}\right)}{\underset{i=1}{\overset{k}{\sum}}\;\sigma_{i}^{2}}\rightarrow0\;;\;\underset{k\rightarrow\infty}{\lim}\frac{\max\left(\sigma_{i}^{2}\upsilon_{i}\right)}{\underset{i=1}{\overset{k}{\sum}}\;\sigma_{i}^{2}\upsilon_{i}}\rightarrow0\right]\quad\quad\right\} 
\end{eqnarray*}
First, using Chebychev's version of the law of large numbers (Loeve
1977) when all the valuations have the same finite expected value
gives,
\[
E\left[\underset{k\rightarrow\infty}{\lim}\frac{1}{\left(k\right)}\frac{\underset{i=1}{\overset{k}{\sum}}\;\underset{j\neq i}{\overset{k}{\sum}}\;\sigma_{j}^{2}\upsilon_{i}}{\underset{i=1}{\overset{k}{\sum}}\sigma_{i}^{2}}\right]=E\left[\underset{k\rightarrow\infty}{\lim}\frac{1}{\left(k\right)}\left\{ \underset{i=1}{\overset{k}{\sum}}\left(\upsilon_{i}-E\left[\upsilon_{i}\right]+E\left[\upsilon_{i}\right]\right)-\frac{\underset{i=1}{\overset{k}{\sum}}\;\sigma_{i}^{2}\upsilon_{i}}{\underset{i=1}{\overset{k}{\sum}}\sigma_{i}^{2}}\right\} \right]
\]
\[
=E\left[\underset{k\rightarrow\infty}{\lim}\frac{1}{\left(k\right)}\left\{ \underset{i=1}{\overset{k}{\sum}}\left(\upsilon_{i}-E\left[\upsilon_{i}\right]\right)+\underset{i=1}{\overset{k}{\sum}}E\left[\upsilon_{i}\right]-\frac{\underset{i=1}{\overset{k}{\sum}}\;\sigma_{i}^{2}\upsilon_{i}}{\underset{i=1}{\overset{k}{\sum}}\sigma_{i}^{2}}\right\} \right]
\]
\[
\Rightarrow E\left[\underset{k\rightarrow\infty}{\lim}\frac{1}{\left(k\right)}\frac{\underset{i=1}{\overset{k}{\sum}}\;\underset{j\neq i}{\overset{k}{\sum}}\;\sigma_{j}^{2}\upsilon_{i}}{\underset{i=1}{\overset{k}{\sum}}\sigma_{i}^{2}}\right]\overset{p}{\longrightarrow}E\left[\upsilon\right]\quad
\]
\[
\left\{ \because\underset{k\rightarrow\infty}{\lim}\frac{1}{\left(k\right)}\left\{ \underset{i=1}{\overset{k}{\sum}}\left(\upsilon_{i}-E\left[\upsilon_{i}\right]\right)\right\} \overset{p}{\longrightarrow}0,\text{ and }\underset{k\rightarrow\infty}{\lim}\frac{1}{\left(k\right)}\underset{i=1}{\overset{k}{\sum}}E\left[\upsilon_{i}\right]=\frac{k}{k}E\left[\upsilon\right]=E\left[\upsilon\right]\right\} 
\]
Second, when the valuations have finite but different means we use
Chebychev's version of the law of large numbers which gives the result,
\[
E\left[\underset{k\rightarrow\infty}{\lim}\frac{1}{\left(k\right)}\frac{\underset{i=1}{\overset{k}{\sum}}\;\underset{j\neq i}{\overset{k}{\sum}}\;\sigma_{j}^{2}\upsilon_{i}}{\underset{i=1}{\overset{k}{\sum}}\sigma_{i}^{2}}\right]=E\left[\underset{k\rightarrow\infty}{\lim}\frac{1}{\left(k\right)}\left\{ \underset{i=1}{\overset{k}{\sum}}\left(\upsilon_{i}\right)-\frac{\underset{i=1}{\overset{k}{\sum}}\;\sigma_{i}^{2}\upsilon_{i}}{\underset{i=1}{\overset{k}{\sum}}\sigma_{i}^{2}}\right\} -E\left[\bar{\upsilon}_{k}\right]+E\left[\bar{\upsilon}_{k}\right]\right]
\]
Here, $\ensuremath{\bar{\upsilon}_{k}=\frac{1}{\left(k\right)}\left\{ \underset{i=1}{\overset{k}{\sum}}\left(\upsilon_{i}\right)\right\} }$
\[
\Rightarrow E\left[\underset{k\rightarrow\infty}{\lim}\frac{1}{\left(k\right)}\frac{\underset{i=1}{\overset{k}{\sum}}\;\underset{j\neq i}{\overset{k}{\sum}}\;\sigma_{j}^{2}\upsilon_{i}}{\underset{i=1}{\overset{k}{\sum}}\sigma_{i}^{2}}\right]\overset{p}{\longrightarrow}E\left[\bar{\upsilon}_{k}\right]\quad
\]
\[
\left\{ \because\underset{k\rightarrow\infty}{\lim}\frac{1}{\left(k\right)}\left\{ \underset{i=1}{\overset{k}{\sum}}\left(\upsilon_{i}\right)\right\} -E\left(\bar{\upsilon}_{k}\right)=\underset{k\rightarrow\infty}{\lim}\bar{\upsilon}_{k}-E\left(\bar{\upsilon}_{k}\right)\overset{p}{\longrightarrow}0\right\} 
\]
\end{proof}

\subsubsection{\label{subsec:Asymptotic-Variance-of}Asymptotic Variance of the
Alternative Weighting Scheme}
\begin{proof}
Looking at the variance of the alternative variance weighting combination
with the limit, $k\rightarrow\infty$, gives,
\[
Var\left[\underset{k\rightarrow\infty}{\lim}\frac{1}{\left(k\right)}\frac{\underset{i=1}{\overset{k}{\sum}}\;\underset{j\neq i}{\overset{k}{\sum}}\;\sigma_{j}^{2}\upsilon_{i}}{\underset{i=1}{\overset{k}{\sum}}\sigma_{i}^{2}}\right]=\underset{k\rightarrow\infty}{\lim}\frac{1}{\left(k^{2}\right)}\frac{\underset{i=1}{\overset{k}{\sum}}\;\left(\underset{j\neq i}{\overset{k}{\sum}}\;\sigma_{j}^{2}\right)^{2}Var\left(\upsilon_{i}\right)}{\left(\underset{i=1}{\overset{k}{\sum}}\sigma_{i}^{2}\right)^{2}}
\]
\[
=\underset{k\rightarrow\infty}{\lim}\frac{1}{\left(k^{2}\right)}\frac{\underset{i=1}{\overset{k}{\sum}}\;\left(\underset{j\neq i}{\overset{k}{\sum}}\;\sigma_{j}^{2}\right)^{2}\sigma_{i}^{2}}{\left(\underset{i=1}{\overset{k}{\sum}}\sigma_{i}^{2}\right)^{2}}
\]
\[
=\underset{k\rightarrow\infty}{\lim}\frac{1}{\left(k^{2}\right)}\frac{\underset{i=1}{\overset{k}{\sum}}\;\left(\underset{j=1}{\overset{k}{\sum}}\;\sigma_{j}^{2}-\sigma_{i}^{2}\right)^{2}\sigma_{i}^{2}}{\left(\underset{i=1}{\overset{k}{\sum}}\sigma_{i}^{2}\right)^{2}}
\]
\[
=\underset{k\rightarrow\infty}{\lim}\frac{1}{\left(k^{2}\right)}\frac{\underset{i=1}{\overset{k}{\sum}}\;\left[\left(\underset{j=1}{\overset{k}{\sum}}\;\sigma_{j}^{2}\right)^{2}+\sigma_{i}^{4}-2\left(\sigma_{i}^{2}\right)\left(\underset{j=1}{\overset{k}{\sum}}\;\sigma_{j}^{2}\right)\right]\sigma_{i}^{2}}{\left(\underset{i=1}{\overset{k}{\sum}}\sigma_{i}^{2}\right)^{2}}
\]
\[
=\underset{k\rightarrow\infty}{\lim}\frac{1}{\left(k^{2}\right)}\left\{ \underset{i=1}{\overset{k}{\sum}}\;\sigma_{i}^{2}\right\} +\frac{1}{\left(k^{2}\right)}\frac{\underset{i=1}{\overset{k}{\sum}}\;\sigma_{i}^{6}}{\left(\underset{i=1}{\overset{k}{\sum}}\sigma_{i}^{2}\right)^{2}}-\frac{1}{\left(k^{2}\right)}\frac{2\left(\underset{i=1}{\overset{k}{\sum}}\;\sigma_{i}^{4}\right)}{\left(\underset{i=1}{\overset{k}{\sum}}\sigma_{i}^{2}\right)}
\]
\[
=0
\]
\[
\left\{ \text{ \ensuremath{\because} }\frac{\left(\underset{i=1}{\overset{k}{\sum}}\;\sigma_{i}^{6}\right)}{\left(\underset{i=1}{\overset{k}{\sum}}\sigma_{i}^{2}\right)^{2}}\leq N_{1};\frac{\left(\underset{i=1}{\overset{k}{\sum}}\;\sigma_{i}^{4}\right)}{\left(\underset{i=1}{\overset{k}{\sum}}\sigma_{i}^{2}\right)}\leq N_{2},\text{ any real numbers and we have finite fourth and sixth moments }\right.
\]
\[
\left.\text{ and since }\underset{k\rightarrow\infty}{\lim}\frac{1}{\left(k^{2}\right)}\underset{i=1}{\overset{k}{\sum}}\sigma_{i}^{2}=0.\vphantom{\left[\frac{\left(\underset{i=1}{\overset{k}{\sum}}\;\sigma_{i}^{6}\right)}{\left(\underset{i=1}{\overset{k}{\sum}}\sigma_{i}^{2}\right)^{2}}\right]}\right\} 
\]
\end{proof}

\subsubsection{\label{subsec:Asymptotic-Expected-Value-Min-Variance}Asymptotic
Expected Value of the Minimum Variance Weighting }
\begin{proof}
We can apply Theorem 1 from (Etemadi 2006) which gives the result
that any weighted average of random variables converges to the simple
average when the weights are monotonically decreasing, which can be
easily satisfied here by construction since we can arrange the weights
based on the variances in a suitable manner. This theorem can also
be used in section \ref{subsec:Asymptotic-Expected-Value}, but the
proof we have provided using the law of large numbers is more direct.
The following common conditions apply,
\[
\left\{ \quad\quad\vphantom{\left[\underset{k\rightarrow\infty}{\lim}\frac{\max\left(\sigma_{i}^{2}\right)}{\underset{i=1}{\overset{k}{\sum}}\;\sigma_{i}^{2}}\rightarrow0,\right]}\text{ Each of the variances and valuations are finite, }\right.
\]
\begin{eqnarray*}
\text{ and the following combination of valuations and variances is uniformly bounded, that is, }
\end{eqnarray*}
\[
\frac{\underset{i=1}{\overset{k}{\sum}}\frac{\upsilon_{i}}{\sigma_{i}^{2}}}{\underset{i=1}{\overset{k}{\sum}}\frac{1}{\sigma_{i}^{2}}}\leq M_{2},\text{ any real number }
\]
\begin{eqnarray*}
\text{ and the covariance between the valuations is zero, }cov\left(v_{i},v_{i}\right)=0;\frac{(\sum_{i=1}^{k}\sigma_{i}^{2})}{k^{2}}\rightarrow0\text{ as }k\rightarrow\infty
\end{eqnarray*}
\begin{eqnarray*}
\left.\text{ and no single term dominates the sum, expressed as, }\left[\underset{k\rightarrow\infty}{\lim}\frac{\max\left(\frac{1}{\sigma_{i}^{2}}\right)}{\underset{i=1}{\overset{k}{\sum}}\;\frac{1}{\sigma_{i}^{2}}}\rightarrow0\;;\;\underset{k\rightarrow\infty}{\lim}\frac{\max\left(\frac{\upsilon_{i}}{\sigma_{i}^{2}}\right)}{\underset{i=1}{\overset{k}{\sum}}\;\frac{\upsilon_{i}}{\sigma_{i}^{2}}}\rightarrow0\right]\quad\quad\right\} 
\end{eqnarray*}
When the valuations have the same expected value, 
\[
E\left[\underset{k\rightarrow\infty}{\lim}\frac{\underset{i=1}{\overset{k}{\sum}}\;\left(\frac{1}{\sigma_{i}^{2}}\upsilon_{i}\right)}{\underset{i=1}{\overset{k}{\sum}}\frac{1}{\sigma_{i}^{2}}}\right]=E\left[\underset{k\rightarrow\infty}{\lim}\frac{\underset{i=1}{\overset{k}{\sum}}\;\frac{1}{\sigma_{i}^{2}}\left(\upsilon_{i}-E\left[\upsilon_{i}\right]+E\left[\upsilon_{i}\right]\right)}{\underset{i=1}{\overset{k}{\sum}}\frac{1}{\sigma_{i}^{2}}}\right]
\]
\[
=E\left[\underset{k\rightarrow\infty}{\lim}\frac{\underset{i=1}{\overset{k}{\sum}}\;\frac{1}{\sigma_{i}^{2}}\left(\upsilon_{i}-E\left[\upsilon_{i}\right]\right)}{\underset{i=1}{\overset{k}{\sum}}\frac{1}{\sigma_{i}^{2}}}+\frac{\underset{i=1}{\overset{k}{\sum}}\;\frac{1}{\sigma_{i}^{2}}E\left[\upsilon_{i}\right]}{\underset{i=1}{\overset{k}{\sum}}\frac{1}{\sigma_{i}^{2}}}\right]
\]
\[
=E\left[\underset{k\rightarrow\infty}{\lim}\frac{\underset{i=1}{\overset{k}{\sum}}\;\frac{1}{\sigma_{i}^{2}}\left(\upsilon_{i}-E\left[\upsilon_{i}\right]\right)}{\underset{i=1}{\overset{k}{\sum}}\frac{1}{\sigma_{i}^{2}}}+\frac{E\left[\upsilon\right]\underset{i=1}{\overset{k}{\sum}}\;\frac{1}{\sigma_{i}^{2}}}{\underset{i=1}{\overset{k}{\sum}}\frac{1}{\sigma_{i}^{2}}}\right]
\]
\[
\overset{p}{\longrightarrow}E\left[\upsilon\right]\quad
\]
\[
\left\{ \because\underset{k\rightarrow\infty}{\lim}\frac{1}{\left(k\right)}\left\{ \underset{i=1}{\overset{k}{\sum}}\left(\upsilon_{i}-E\left[\upsilon_{i}\right]\right)\right\} \overset{p}{\longrightarrow}0\Longrightarrow\underset{k\rightarrow\infty}{\lim}\frac{\underset{i=1}{\overset{k}{\sum}}\;\frac{1}{\sigma_{i}^{2}}\left(\upsilon_{i}-E\left[\upsilon_{i}\right]\right)}{\underset{i=1}{\overset{k}{\sum}}\frac{1}{\sigma_{i}^{2}}}\overset{p}{\longrightarrow}0\right\} 
\]
Similarly when the valuations have different means we get, 
\[
E\left[\underset{k\rightarrow\infty}{\lim}\frac{\underset{i=1}{\overset{k}{\sum}}\;\left(\frac{1}{\sigma_{i}^{2}}\upsilon_{i}\right)}{\underset{i=1}{\overset{k}{\sum}}\frac{1}{\sigma_{i}^{2}}}\right]=E\left[\underset{k\rightarrow\infty}{\lim}\frac{\underset{i=1}{\overset{k}{\sum}}\;\frac{1}{\sigma_{i}^{2}}\left(\upsilon_{i}\right)}{\underset{i=1}{\overset{k}{\sum}}\frac{1}{\sigma_{i}^{2}}}-E\left[\bar{\upsilon}_{k}\right]+E\left[\bar{\upsilon}_{k}\right]\right]
\]
\[
\overset{p}{\longrightarrow}E\left[\bar{\upsilon}_{k}\right]\quad
\]
Here, $\ensuremath{\bar{\upsilon}_{k}=\frac{1}{\left(k\right)}\left\{ \underset{i=1}{\overset{k}{\sum}}\left(\upsilon_{i}\right)\right\} }$
\[
\left\{ \because\underset{k\rightarrow\infty}{\lim}\frac{1}{\left(k\right)}\left\{ \underset{i=1}{\overset{k}{\sum}}\left(\upsilon_{i}\right)\right\} -E\left(\bar{\upsilon}_{k}\right)=\underset{k\rightarrow\infty}{\lim}\bar{\upsilon}_{k}-E\left(\bar{\upsilon}_{k}\right)\overset{p}{\longrightarrow}0\Longrightarrow\underset{k\rightarrow\infty}{\lim}\frac{\underset{i=1}{\overset{k}{\sum}}\;\frac{1}{\sigma_{i}^{2}}\left(\upsilon_{i}\right)}{\underset{i=1}{\overset{k}{\sum}}\frac{1}{\sigma_{i}^{2}}}-E\left[\bar{\upsilon}_{k}\right]\overset{p}{\longrightarrow}0\right\} 
\]
Alternately, we can show the convergence of the minimum variance scheme
by considering the ratio (End-note: \ref{Limit-Comparison-Test},
Limit Comparison Test) of the $k^{th}$ terms of the two weighting
schemes, 
\[
\underset{k\rightarrow\infty}{\lim}\left\{ \frac{1}{\left(k\right)}\frac{\underset{j\neq i}{\overset{k}{\sum}}\;\sigma_{j}^{2}\upsilon_{i}}{\underset{j=1}{\overset{k}{\sum}}\sigma_{j}^{2}}\right\} /\left\{ \frac{\left(\frac{1}{\sigma_{i}^{2}}\upsilon_{i}\right)}{\underset{j=1}{\overset{k}{\sum}}\frac{1}{\sigma_{j}^{2}}}\right\} =\underset{k\rightarrow\infty}{\lim}\frac{1}{\left(k\right)}\frac{\left(\underset{j\neq i}{\overset{k}{\sum}}\;\sigma_{j}^{2}\upsilon_{i}\right)}{\left(\underset{j=1}{\overset{k}{\sum}}\sigma_{j}^{2}\right)}\frac{\left(\underset{j=1}{\overset{k}{\sum}}\frac{1}{\sigma_{j}^{2}}\right)}{\left(\frac{1}{\sigma_{i}^{2}}\upsilon_{i}\right)}
\]
Without loss of generality we can set $\upsilon_{i}=\upsilon_{k}$
and modify the expression as follows,
\[
=\underset{k\rightarrow\infty}{\lim}\frac{1}{\left(k\right)}\frac{\left(\underset{j=1}{\overset{k-1}{\sum}}\;\sigma_{j}^{2}\right)}{\left(\underset{j=1}{\overset{k}{\sum}}\sigma_{j}^{2}\right)}\frac{\left(\underset{j=1}{\overset{k}{\sum}}\frac{1}{\sigma_{j}^{2}}\right)}{\left(\frac{1}{\sigma_{k}^{2}}\right)}
\]
\[
=\underset{k\rightarrow\infty}{\lim}\frac{1}{\left(k\right)}\frac{\left(\underset{j=1}{\overset{k-1}{\sum}}\;\sigma_{j}^{2}\right)}{\left(\underset{j=1}{\overset{k}{\sum}}\sigma_{j}^{2}\right)}\frac{\left(\frac{1}{\sigma_{k}^{2}}+\underset{j=1}{\overset{k-1}{\sum}}\frac{1}{\sigma_{j}^{2}}\right)}{\left(\frac{1}{\sigma_{k}^{2}}\right)}
\]
\[
=\underset{k\rightarrow\infty}{\lim}\frac{1}{\left(k\right)}\frac{\left(\underset{j=1}{\overset{k-1}{\sum}}\;\sigma_{j}^{2}\right)}{\left(\underset{j=1}{\overset{k}{\sum}}\sigma_{j}^{2}\right)}+\frac{\sigma_{k}^{2}}{\left(k\right)}\frac{1}{\left(\underset{j=1}{\overset{k}{\sum}}\sigma_{j}^{2}\right)}\left(\underset{j=1}{\overset{k-1}{\sum}}\;\sigma_{j}^{2}\right)\left(\underset{j=1}{\overset{k-1}{\sum}}\frac{1}{\sigma_{j}^{2}}\right)
\]
\[
\geq\underset{k\rightarrow\infty}{\lim}\frac{1}{\left(k\right)}\frac{\left(\underset{j=1}{\overset{k-1}{\sum}}\;\sigma_{j}^{2}\right)}{\left(\underset{j=1}{\overset{k}{\sum}}\sigma_{j}^{2}\right)}+\frac{\sigma_{k}^{2}}{\left(k\right)}\frac{\left(k-1\right)^{2}}{\left(\underset{j=1}{\overset{k}{\sum}}\sigma_{j}^{2}\right)}\left\{ \because\text{ Cauchy Schwartz Inequality gives }\left(\underset{j=1}{\overset{k-1}{\sum}}\;\sigma_{j}^{2}\right)\left(\underset{j=1}{\overset{k-1}{\sum}}\frac{1}{\sigma_{j}^{2}}\right)\geq\left(k-1\right)^{2}\right\} 
\]
\[
=\underset{k\rightarrow\infty}{\lim}\frac{1}{\left(k\right)}\frac{1}{\left(1+\sigma_{k}^{2}/\left\{ \underset{j=1}{\overset{k-1}{\sum}}\sigma_{j}^{2}\right\} \right)}+\frac{\left(k^{2}+1-2k\right)\sigma_{k}^{2}}{\left(k\right)}\frac{1}{\left(\underset{j=1}{\overset{k}{\sum}}\sigma_{j}^{2}\right)}
\]
\[
=\underset{k\rightarrow\infty}{\lim}\frac{1}{\left(k\right)}\frac{1}{\left(1+\sigma_{k}^{2}/\left\{ \underset{j=1}{\overset{k-1}{\sum}}\sigma_{j}^{2}\right\} \right)}+\left(k+\frac{1}{k}-2\right)\sigma_{k}^{2}\frac{1}{\left(\underset{j=1}{\overset{k}{\sum}}\sigma_{j}^{2}\right)}
\]
\[
=\underset{k\rightarrow\infty}{\lim}\frac{1}{\left(k\right)}\frac{1}{\left(1+\sigma_{k}^{2}/\left\{ \underset{j=1}{\overset{k-1}{\sum}}\sigma_{j}^{2}\right\} \right)}+\sigma_{k}^{2}\frac{1}{\frac{1}{k}\left(\underset{j=1}{\overset{k}{\sum}}\sigma_{j}^{2}\right)}+\left(\frac{1}{k}-2\right)\sigma_{k}^{2}\frac{1}{\left(\underset{j=1}{\overset{k}{\sum}}\sigma_{j}^{2}\right)}
\]
\[
=\frac{\sigma_{k}^{2}}{\bar{\sigma}^{2}}>0\qquad\text{ Here the average variance is, }\bar{\sigma}^{2}=\frac{1}{\left(k\right)}\left(\underset{j=1}{\overset{k}{\sum}}\sigma_{j}^{2}\right)
\]
Alternately we can write the ratios as,
\[
\underset{k\rightarrow\infty}{\lim}\left\{ \frac{1}{\left(k\right)}\frac{\underset{j\neq i}{\overset{k}{\sum}}\;\sigma_{j}^{2}\upsilon_{i}}{\underset{j=1}{\overset{k}{\sum}}\sigma_{j}^{2}}\right\} /\left\{ \frac{\left(\frac{1}{\sigma_{i}^{2}}\upsilon_{i}\right)}{\underset{j=1}{\overset{k}{\sum}}\frac{1}{\sigma_{j}^{2}}}\right\} =\underset{k\rightarrow\infty}{\lim}\frac{1}{\left(k\right)}\frac{\left(\underset{j=1}{\overset{k-1}{\sum}}\;\sigma_{j}^{2}\right)}{\left(\underset{j=1}{\overset{k}{\sum}}\sigma_{j}^{2}\right)}+\frac{\sigma_{k}^{2}}{\left(k\right)}\frac{1}{\left(\underset{j=1}{\overset{k}{\sum}}\sigma_{j}^{2}\right)}\left(\underset{j=1}{\overset{k-1}{\sum}}\;\sigma_{j}^{2}\right)\left(\underset{j=1}{\overset{k-1}{\sum}}\frac{1}{\sigma_{j}^{2}}\right)
\]
\[
=\underset{k\rightarrow\infty}{\lim}\frac{1}{\left(k\right)}\frac{1}{\left(1+\sigma_{k}^{2}/\left\{ \underset{j=1}{\overset{k-1}{\sum}}\sigma_{j}^{2}\right\} \right)}+\frac{\sigma_{k}^{2}}{\left(1+\sigma_{k}^{2}/\left\{ \underset{j=1}{\overset{k-1}{\sum}}\sigma_{j}^{2}\right\} \right)}\frac{\left(\underset{j=1}{\overset{k-1}{\sum}}\frac{1}{\sigma_{j}^{2}}\right)}{\left(k\right)}
\]
\[
=\frac{\sigma_{k}^{2}}{\tilde{\sigma}^{2}}<\infty\qquad\text{ Here the average of the reciprocal variance is, }\frac{1}{\tilde{\sigma}^{2}}=\frac{1}{\left(k\right)}\left(\underset{j=1}{\overset{k-1}{\sum}}\frac{1}{\sigma_{j}^{2}}\right)
\]
The ratio of the $k^{th}$ terms of the two weighting schemes is greater
than $\frac{\sigma_{k}^{2}}{\bar{\sigma}^{2}}>0$ and is equal to
$\frac{\sigma_{k}^{2}}{\tilde{\sigma}^{2}}<\infty$. Hence both the
series converge.
\end{proof}

\subsubsection{\label{subsec:Rate-of-Convergence}Rate of Convergence of Alternative
Weighting and Minimum Variance Weighting Schemes}
\begin{proof}
We next arrange terms in each weighting scheme so that the successive
terms are decreasing. This means that the sequence (successive terms
when there are infinitely many terms) has to converge to zero for
the series (sum of terms when there are infinitely many terms) to
converge, which we have already shown. We apply the ratio test (End-note
\ref{Ratio-Test}, Ratio Test) for the convergence of a sequence.
First, we consider the minimum variance weighting,
\[
\underset{k\rightarrow\infty}{\lim}\left\{ \frac{\left(\frac{1}{\sigma_{k+1}^{2}}\upsilon_{k+1}\right)}{\underset{j=1}{\overset{k+1}{\sum}}\frac{1}{\sigma_{j}^{2}}}\right\} /\left\{ \frac{\left(\frac{1}{\sigma_{k}^{2}}\upsilon_{k}\right)}{\underset{j=1}{\overset{k}{\sum}}\frac{1}{\sigma_{j}^{2}}}\right\} <1
\]
\[
\Rightarrow\underset{k\rightarrow\infty}{\lim}\left\{ \frac{\left(\frac{\upsilon_{k+1}}{\sigma_{k+1}^{2}}\right)\underset{j=1}{\overset{k}{\sum}}\frac{1}{\sigma_{j}^{2}}}{\left(\frac{\upsilon_{k}}{\sigma_{k}^{2}}\right)\underset{j=1}{\overset{k+1}{\sum}}\frac{1}{\sigma_{j}^{2}}}\right\} <1
\]
\[
\Rightarrow\underset{k\rightarrow\infty}{\lim}\left\{ \left(\frac{\upsilon_{k+1}}{\upsilon_{k}}\right)\left(\frac{\sigma_{k}^{2}}{\sigma_{k+1}^{2}}\right)\frac{1}{\left[1+\frac{1}{\sigma_{K+1}^{2}}/\left(\underset{j=1}{\overset{k+1}{\sum}}\frac{1}{\sigma_{j}^{2}}\right)\right]}\right\} <1
\]
\[
\Rightarrow\left(\frac{\upsilon_{k+1}}{\upsilon_{k}}\right)\left(\frac{\sigma_{k}^{2}}{\sigma_{k+1}^{2}}\right)<1\quad\left\{ \because\text{ No single term dominates the sum, expressed as, }\underset{k\rightarrow\infty}{\lim}\frac{\max\left(\frac{1}{\sigma_{i}^{2}}\right)}{\underset{i=1}{\overset{k}{\sum}}\;\frac{1}{\sigma_{i}^{2}}}\rightarrow0\;\right\} 
\]
Second we consider the alternative weighting scheme,
\[
\underset{k\rightarrow\infty}{\lim}\left\{ \frac{1}{\left(k+1\right)}\frac{\underset{j=1}{\overset{k}{\sum}}\;\sigma_{j}^{2}\upsilon_{k+1}}{\underset{j=1}{\overset{k+1}{\sum}}\sigma_{j}^{2}}\right\} /\left\{ \frac{1}{\left(k\right)}\frac{\underset{j=1}{\overset{k-1}{\sum}}\;\sigma_{j}^{2}\upsilon_{k}}{\underset{j=1}{\overset{k}{\sum}}\sigma_{j}^{2}}\right\} <1
\]
\[
\Rightarrow\underset{k\rightarrow\infty}{\lim}\left\{ \left(\frac{\upsilon_{k+1}}{\upsilon_{k}}\right)\left(\frac{k}{k+1}\right)\frac{\underset{j=1}{\overset{k}{\sum}}\;\sigma_{j}^{2}}{\underset{j=1}{\overset{k+1}{\sum}}\sigma_{j}^{2}}\right\} \left\{ \frac{\underset{j=1}{\overset{k}{\sum}}\sigma_{j}^{2}}{\underset{j=1}{\overset{k-1}{\sum}}\;\sigma_{j}^{2}}\right\} <1
\]
\[
\Rightarrow\underset{k\rightarrow\infty}{\lim}\left\{ \left(\frac{\upsilon_{k+1}}{\upsilon_{k}}\right)\left(\frac{1}{1+\frac{1}{k}}\right)\frac{\left[1+\sigma_{k}^{2}/\left(\underset{j=1}{\overset{k-1}{\sum}}\sigma_{j}^{2}\right)\right]}{\left[1+\sigma_{k+1}^{2}/\left(\underset{j=1}{\overset{k}{\sum}}\sigma_{j}^{2}\right)\right]}\right\} <1
\]
\[
\Rightarrow\left(\frac{\upsilon_{k+1}}{\upsilon_{k}}\right)<1\quad\left\{ \because\text{ No single term dominates the sum, expressed as, }\underset{k\rightarrow\infty}{\lim}\frac{\max\left(\sigma_{i}^{2}\right)}{\underset{i=1}{\overset{k}{\sum}}\sigma_{i}^{2}}\rightarrow0\;\right\} 
\]
We then compare the rate of convergence (End-note \ref{Rate-Convergence},
Rate of Convergence) of the two sequences as follows, 
\[
\underset{k\rightarrow\infty}{\lim}\left\{ \frac{\left(\frac{1}{\sigma_{k+1}^{2}}\upsilon_{k+1}\right)}{\underset{j=1}{\overset{k+1}{\sum}}\frac{1}{\sigma_{j}^{2}}}\right\} /\left\{ \frac{\left(\frac{1}{\sigma_{k}^{2}}\upsilon_{k}\right)}{\underset{j=1}{\overset{k}{\sum}}\frac{1}{\sigma_{j}^{2}}}\right\} /\left\{ \frac{1}{\left(k+1\right)}\frac{\underset{j=1}{\overset{k}{\sum}}\;\sigma_{j}^{2}\upsilon_{k+1}}{\underset{j=1}{\overset{k+1}{\sum}}\sigma_{j}^{2}}\right\} /\left\{ \frac{1}{\left(k\right)}\frac{\underset{j=1}{\overset{k-1}{\sum}}\;\sigma_{j}^{2}\upsilon_{k}}{\underset{j=1}{\overset{k}{\sum}}\sigma_{j}^{2}}\right\} 
\]
\[
=\underset{k\rightarrow\infty}{\lim}\left\{ \frac{\left(\frac{1}{\sigma_{k+1}^{2}}\upsilon_{k+1}\right)}{\underset{j=1}{\overset{k+1}{\sum}}\frac{1}{\sigma_{j}^{2}}}\right\} \left\{ \frac{\underset{j=1}{\overset{k}{\sum}}\frac{1}{\sigma_{j}^{2}}}{\left(\frac{1}{\sigma_{k}^{2}}\upsilon_{k}\right)}\right\} /\left\{ \frac{k}{\left(k+1\right)}\frac{\underset{j=1}{\overset{k}{\sum}}\;\sigma_{j}^{2}\upsilon_{k+1}}{\underset{j=1}{\overset{k+1}{\sum}}\sigma_{j}^{2}}\right\} \left\{ \frac{\underset{j=1}{\overset{k}{\sum}}\sigma_{j}^{2}}{\underset{j=1}{\overset{k-1}{\sum}}\;\sigma_{j}^{2}\upsilon_{k}}\right\} 
\]
\[
=\underset{k\rightarrow\infty}{\lim}\left\{ \frac{\left(\frac{1}{\sigma_{k+1}^{2}}\upsilon_{k+1}\right)}{\underset{j=1}{\overset{k+1}{\sum}}\frac{1}{\sigma_{j}^{2}}}\right\} \left\{ \frac{\underset{j=1}{\overset{k}{\sum}}\frac{1}{\sigma_{j}^{2}}}{\left(\frac{1}{\sigma_{k}^{2}}\upsilon_{k}\right)}\right\} \left\{ \frac{\left(k+1\right)}{\left(k\right)}\frac{\underset{j=1}{\overset{k+1}{\sum}}\sigma_{j}^{2}}{\underset{j=1}{\overset{k}{\sum}}\;\sigma_{j}^{2}\upsilon_{k+1}}\right\} \left\{ \frac{\underset{j=1}{\overset{k-1}{\sum}}\;\sigma_{j}^{2}\upsilon_{k}}{\underset{j=1}{\overset{k}{\sum}}\sigma_{j}^{2}}\right\} 
\]
\[
=\underset{k\rightarrow\infty}{\lim}\left(\frac{\underset{j=1}{\overset{k+1}{\sum}}\sigma_{j}^{2}}{\underset{j=1}{\overset{k+1}{\sum}}\frac{1}{\sigma_{j}^{2}}}\right)\left(\frac{\underset{j=1}{\overset{k}{\sum}}\frac{1}{\sigma_{j}^{2}}}{\underset{j=1}{\overset{k}{\sum}}\;\sigma_{j}^{2}}\right)\left(\frac{\sigma_{k}^{2}}{\sigma_{k+1}^{2}}\right)\left(1+\frac{1}{k}\right)\left\{ \frac{\underset{j=1}{\overset{k-1}{\sum}}\;\sigma_{j}^{2}}{\underset{j=1}{\overset{k}{\sum}}\sigma_{j}^{2}}\right\} 
\]
We use the following result, End-note \ref{Product of Sum and their Reciprocals}:
$\left(\underset{j=1}{\overset{k}{\sum}}\;\sigma_{j}^{2}\right)\left(\underset{j=1}{\overset{k}{\sum}}\frac{1}{\sigma_{j}^{2}}\right)\geq k^{2}$,
about the product of sums of $k$ numbers and their reciprocals is
greater than $k^{2}$; which follows easily from the Cauchy Schwartz
Inequality, helping us to simplify as follows, 
\[
\geq\underset{k\rightarrow\infty}{\lim}\left\{ \frac{\left(\underset{j=1}{\overset{k+1}{\sum}}\sigma_{j}^{2}\right)^{2}}{\left(k+1\right)^{2}}\right\} \left\{ \frac{k^{2}}{\left(\underset{j=1}{\overset{k}{\sum}}\;\sigma_{j}^{2}\right)^{2}}\right\} \left(\frac{\sigma_{k}^{2}}{\sigma_{k+1}^{2}}\right)\left(1+\frac{1}{k}\right)\left\{ \frac{\underset{j=1}{\overset{k-1}{\sum}}\;\sigma_{j}^{2}}{\underset{j=1}{\overset{k}{\sum}}\sigma_{j}^{2}}\right\} 
\]
\[
=\underset{k\rightarrow\infty}{\lim}\left\{ \frac{\left(\underset{j=1}{\overset{k+1}{\sum}}\sigma_{j}^{2}\right)}{\left(\underset{j=1}{\overset{k}{\sum}}\;\sigma_{j}^{2}\right)}\right\} ^{2}\left(\frac{\sigma_{k}^{2}}{\sigma_{k+1}^{2}}\right)\left(\frac{k}{k+1}\right)\left\{ \frac{\underset{j=1}{\overset{k-1}{\sum}}\;\sigma_{j}^{2}}{\underset{j=1}{\overset{k}{\sum}}\sigma_{j}^{2}}\right\} 
\]
\[
=\underset{k\rightarrow\infty}{\lim}\left\{ 1+\frac{\sigma_{k+1}^{2}}{\left(\underset{j=1}{\overset{k}{\sum}}\;\sigma_{j}^{2}\right)}\right\} ^{2}\left(\frac{\sigma_{k}^{2}}{\sigma_{k+1}^{2}}\right)\left(\frac{k}{k+1}\right)\left\{ \frac{1}{1+\sigma_{k}^{2}/\left(\underset{j=1}{\overset{k-1}{\sum}}\sigma_{j}^{2}\right)}\right\} 
\]
\[
=\left(\frac{\sigma_{k}^{2}}{\sigma_{k+1}^{2}}\right)
\]
Combining the convergence criterion for the two sequences with the
above ratio of the rates of convergence shows that the alternative
weighting scheme converges faster than the minimum variance scheme
as long as following conditions hold,
\[
\left(\frac{\sigma_{k}^{2}}{\sigma_{k+1}^{2}}\right)>1\;;\;\left(\frac{\upsilon_{k+1}}{\upsilon_{k}}\right)<1\;;\;\left(\frac{\upsilon_{k+1}}{\upsilon_{k}}\right)\left(\frac{\sigma_{k}^{2}}{\sigma_{k+1}^{2}}\right)<1
\]
 
\end{proof}

\subsubsection{\label{subsec:Asymptotic-Variance-Min-Variance}Asymptotic Variance
of the Minimum Variance Combination}
\begin{proof}
Consider,
\[
V\left[\underset{k\rightarrow\infty}{\lim}\frac{\underset{i=1}{\overset{k}{\sum}}\;\left(\frac{1}{\sigma_{i}^{2}}\right)\upsilon_{i}}{\underset{i=1}{\overset{k}{\sum}}\frac{1}{\sigma_{i}^{2}}}\right]=\underset{k\rightarrow\infty}{\lim}\frac{\underset{i=1}{\overset{k}{\sum}}\;\left(\frac{1}{\sigma_{i}^{4}}\right)\sigma_{i}^{2}}{\left(\underset{i=1}{\overset{k}{\sum}}\frac{1}{\sigma_{i}^{2}}\right)^{2}}
\]
\[
=\underset{k\rightarrow\infty}{\lim}\frac{\underset{i=1}{\overset{k}{\sum}}\;\left(\frac{1}{\sigma_{i}^{2}}\right)}{\left(\underset{i=1}{\overset{k}{\sum}}\frac{1}{\sigma_{i}^{2}}\right)^{2}}
\]
\[
=\underset{k\rightarrow\infty}{\lim}\frac{1}{\left(\underset{i=1}{\overset{k}{\sum}}\frac{1}{\sigma_{i}^{2}}\right)}<\infty
\]
\[
=0\quad\text{ if }\underset{k\rightarrow\infty}{\lim}\left(\underset{i=1}{\overset{k}{\sum}}\frac{1}{\sigma_{i}^{2}}\right)\rightarrow\infty
\]
\end{proof}

\subsubsection{\label{subsec:Minimum-Variance-Weighting}Minimum Variance Weighting
as Number of Valuations Increases}
\begin{proof}
We show by induction that the minimum variance weighting holds for
any number of valuations when the covariance terms are zero. Let us
first consider the case of only two valuations, $\upsilon_{1}$ and
$\upsilon_{2}$. In this case, $k=2$. Let us combine them using the
relative weights $w_{1}$ and $w_{2}$ such that $w_{1}+w_{2}=1$.
The variance of the combined valuation in this case, $\sigma_{2c}^{2}$,
is given by,
\[
\sigma_{2c}^{2}=w_{1}^{2}\sigma_{1}^{2}+w_{2}^{2}\sigma_{2}^{2}
\]
\[
\sigma_{2c}^{2}=w_{1}^{2}\sigma_{1}^{2}+\left(1-w_{1}\right)^{2}\sigma_{2}^{2}
\]
Differentiating $\sigma_{2c}^{2}$ with respect to $w_{1}$ and setting
the first order condition, FOC, gives,
\[
2w_{1}\sigma_{1}^{2}-2\left(1-w_{1}\right)\sigma_{2}^{2}=0
\]
\[
w_{1}=\frac{\sigma_{2}^{2}}{\sigma_{1}^{2}+\sigma_{2}^{2}}=\frac{\frac{1}{\sigma_{1}^{2}}}{\frac{1}{\sigma_{1}^{2}}+\frac{1}{\sigma_{2}^{2}}}
\]
\[
w_{2}=\frac{\sigma_{1}^{2}}{\sigma_{1}^{2}+\sigma_{2}^{2}}=\frac{\frac{1}{\sigma_{2}^{2}}}{\frac{1}{\sigma_{1}^{2}}+\frac{1}{\sigma_{2}^{2}}}
\]
The variance of the combinations becomes,
\[
\sigma_{2c}^{2}=\left(\frac{\frac{1}{\sigma_{1}^{2}}}{\frac{1}{\sigma_{1}^{2}}+\frac{1}{\sigma_{2}^{2}}}\right)^{2}\sigma_{1}^{2}+\left(\frac{\frac{1}{\sigma_{2}^{2}}}{\frac{1}{\sigma_{1}^{2}}+\frac{1}{\sigma_{2}^{2}}}\right)^{2}\sigma_{2}^{2}
\]
\[
\sigma_{2c}^{2}=\frac{\frac{1}{\sigma_{1}^{2}}}{\left\{ \frac{1}{\sigma_{1}^{2}}+\frac{1}{\sigma_{2}^{2}}\right\} ^{2}}+\frac{\frac{1}{\sigma_{2}^{2}}}{\left\{ \frac{1}{\sigma_{1}^{2}}+\frac{1}{\sigma_{2}^{2}}\right\} ^{2}}
\]
\[
\sigma_{2c}^{2}=\frac{1}{\left\{ \frac{1}{\sigma_{1}^{2}}+\frac{1}{\sigma_{2}^{2}}\right\} ^{2}}\left\{ \frac{1}{\sigma_{1}^{2}}+\frac{1}{\sigma_{2}^{2}}\right\} 
\]
\[
\sigma_{2c}^{2}=\frac{1}{\left\{ \frac{1}{\sigma_{1}^{2}}+\frac{1}{\sigma_{2}^{2}}\right\} }\Leftrightarrow\frac{1}{\sigma_{2c}^{2}}=\left\{ \frac{1}{\sigma_{1}^{2}}+\frac{1}{\sigma_{2}^{2}}\right\} 
\]
Second order conditions show that this is a minimum point.
\[
\frac{d^{2}\left(\sigma_{2c}^{2}\right)}{d\left(w_{1}\right)^{2}}=2\sigma_{1}^{2}+2\sigma_{2}^{2}>0
\]
Let us assume that these minimum variance weights hold for the case
of $k=K$ valuations. That is,
\[
w_{i}=\frac{\frac{1}{\sigma_{i}^{2}}}{\frac{1}{\sigma_{1}^{2}}+\cdots+\frac{1}{\sigma_{K}^{2}}}\quad;\;i=1,\ldots,K
\]
The variance of the combination when there are $k=K$ valuations,
$\sigma_{Kc}^{2}$, is,
\[
\sigma_{Kc}^{2}=w_{1}^{2}\sigma_{1}^{2}+\cdots+w_{K}^{2}\sigma_{K}^{2}
\]
\[
\sigma_{Kc}^{2}=\frac{\frac{1}{\sigma_{1}^{2}}}{\left\{ \frac{1}{\sigma_{1}^{2}}+\cdots+\frac{1}{\sigma_{K}^{2}}\right\} ^{2}}+\cdots+\frac{\frac{1}{\sigma_{K}^{2}}}{\left\{ \frac{1}{\sigma_{1}^{2}}+\cdots+\frac{1}{\sigma_{K}^{2}}\right\} ^{2}}
\]
\[
\sigma_{Kc}^{2}=\frac{1}{\left\{ \frac{1}{\sigma_{1}^{2}}+\cdots+\frac{1}{\sigma_{K}^{2}}\right\} ^{2}}\left\{ \frac{1}{\sigma_{1}^{2}}+\cdots+\frac{1}{\sigma_{K}^{2}}\right\} 
\]
\[
\sigma_{Kc}^{2}=\frac{1}{\left\{ \frac{1}{\sigma_{1}^{2}}+\cdots+\frac{1}{\sigma_{K}^{2}}\right\} }\Leftrightarrow\frac{1}{\sigma_{Kc}^{2}}=\left\{ \frac{1}{\sigma_{1}^{2}}+\cdots+\frac{1}{\sigma_{K}^{2}}\right\} 
\]
We show by induction that this weighting scheme gives the minimum
variance, $\sigma_{K+1c}^{2}$, for the case $k=K+1$. We find the
optimal weights when a new valuation is combined with the existing
$K$ valuations such that the sum of the weights of this new combination
is one, $w_{Kc}+w_{K+1}=1$.
\[
\sigma_{K+1c}^{2}=w_{Kc}^{2}\sigma_{Kc}^{2}+w_{K+1}^{2}\sigma_{K+1}^{2}
\]
\[
\sigma_{K+1c}^{2}=w_{Kc}^{2}\sigma_{Kc}^{2}+\left(1-w_{Kc}\right)^{2}\sigma_{K+1}^{2}
\]
Differentiating with respect to $w_{Kc}$ and setting the first order
condition, FOC, gives,
\[
2w_{Kc}\sigma_{Kc}^{2}-2\left(1-w_{Kc}\right)\sigma_{K+1}^{2}=0
\]
\[
w_{Kc}=\frac{\sigma_{K+1}^{2}}{\sigma_{Kc}^{2}+\sigma_{K+1}^{2}}=\frac{\frac{1}{\sigma_{Kc}^{2}}}{\frac{1}{\sigma_{Kc}^{2}}+\frac{1}{\sigma_{K+1}^{2}}}
\]
The weight of the previous $K$ valuations then becomes, 
\[
w_{i}=\left\{ \frac{\frac{1}{\sigma_{i}^{2}}}{\frac{1}{\sigma_{1}^{2}}+\cdots+\frac{1}{\sigma_{K}^{2}}}\right\} \left\{ \frac{\frac{1}{\sigma_{Kc}^{2}}}{\frac{1}{\sigma_{Kc}^{2}}+\frac{1}{\sigma_{K+1}^{2}}}\right\} \quad;\;i=1,\ldots,K
\]
\[
w_{i}=\left\{ \frac{\frac{1}{\sigma_{i}^{2}}}{\frac{1}{\sigma_{1}^{2}}+\cdots+\frac{1}{\sigma_{K}^{2}}+\frac{1}{\sigma_{K+1}^{2}}}\right\} \quad;\;i=1,\ldots,K
\]
The weights of the new $\left(K+1\right)^{th}$ valuation is,
\[
w_{K+1}=\frac{\frac{1}{\sigma_{K+1}^{2}}}{\frac{1}{\sigma_{Kc}^{2}}+\frac{1}{\sigma_{K+1}^{2}}}=\left\{ \frac{\frac{1}{\sigma_{K+1}^{2}}}{\frac{1}{\sigma_{1}^{2}}+\cdots+\frac{1}{\sigma_{K}^{2}}+\frac{1}{\sigma_{K+1}^{2}}}\right\} 
\]
The variance of the combination is, 
\[
\sigma_{K+1c}^{2}=w_{Kc}^{2}\sigma_{Kc}^{2}+w_{K+1}^{2}\sigma_{K+1}^{2}
\]
\[
\sigma_{K+1c}^{2}=\left(\frac{\frac{1}{\sigma_{Kc}^{2}}}{\frac{1}{\sigma_{Kc}^{2}}+\frac{1}{\sigma_{K+1}^{2}}}\right)^{2}\sigma_{Kc}^{2}+\left\{ \frac{\frac{1}{\sigma_{K+1}^{2}}}{\frac{1}{\sigma_{1}^{2}}+\cdots+\frac{1}{\sigma_{K}^{2}}+\frac{1}{\sigma_{K+1}^{2}}}\right\} ^{2}\sigma_{K+1}^{2}
\]
\[
\sigma_{K+1c}^{2}=\left(\frac{1}{\sigma_{Kc}^{2}}\right)\frac{1}{\left\{ \frac{1}{\sigma_{1}^{2}}+\cdots+\frac{1}{\sigma_{K}^{2}}+\frac{1}{\sigma_{K+1}^{2}}\right\} ^{2}}+\left(\frac{1}{\sigma_{K+1}^{2}}\right)\frac{1}{\left\{ \frac{1}{\sigma_{1}^{2}}+\cdots+\frac{1}{\sigma_{K}^{2}}+\frac{1}{\sigma_{K+1}^{2}}\right\} ^{2}}
\]
\[
\sigma_{K+1c}^{2}=\frac{1}{\left\{ \frac{1}{\sigma_{1}^{2}}+\cdots+\frac{1}{\sigma_{K}^{2}}+\frac{1}{\sigma_{K+1}^{2}}\right\} }
\]
This completes the proof by induction.
\end{proof}

\section{Appendix: Dictionary of Notation and Terminology for the Auction
Strategy}
\begin{itemize}
\item $x_{i}$, the valuation of intermediary or bidder $i$. This is a
realization of the random variable $X_{i}$ which bidder $i$ and
only bidder $i$ knows for sure.
\item $x_{i}\sim F\left[0,\omega\right]$, $x_{i}$ is symmetric and independently
distributed according to the distribution $F$ over the interval $\left[0,\omega\right]$. 
\item $F,$ is increasing and has full support, which is the non-negative
real line $\left[0,\infty\right]$.
\item $f=F',$ is the continuous density function of $F$.
\item $x_{i}\sim U\left[0,\omega\right]$ when we consider the uniform distribution.
\item $x_{i}\sim LN\left[0,\omega\right]$ when we consider the log normal
distribution.
\item $M,$ is the total number of bidders.
\item $f_{i},\,F_{i}$ , are the continuous density function and distribution
of bidder $i$ in the asymmetric case.
\item $r\geq0$, is the reserve price set by the auction seller.
\item $\beta_{i}:\left[0,\omega\right]\rightarrow\Re_{+}$ is a increasing
function that gives the strategy for bidder $i$. We let $\beta_{i}\left(x_{i}\right)=b_{i}$.
We must have $\beta_{i}\left(0\right)=0$.
\item $\phi_{i}\equiv\beta_{i}^{-1}$ is the inverse of the bidding strategy
$\beta_{i}$. This means, $x_{i}=\beta_{i}^{-1}\left(b_{i}\right)=\phi_{i}\left(b_{i}\right)$.
\item $x_{i}\sim F_{i}\left[0,\omega_{i}\right]$. Here, $x_{i}$ is asymmetric
and is independently distributed according to the distribution $F_{i}$
over the interval $\left[0,\omega_{i}\right]$. 
\item $\beta:\left[0,\omega\right]\rightarrow\Re_{+}$ is the strategy of
all the bidders in a symmetric equilibrium. We let $\beta\left(x\right)=b,\;x$
is the valuation of any bidder. We also have $b\leq\beta\left(x\right)\;\text{and}\;\beta\left(0\right)=0$.
\item $Y_{1}\equiv Y_{1}^{M-1}$ , the random variable that denotes the
highest value, say for bidder 1, among the $M-1$ other bidders.
\item $Y_{1},$ is the highest order statistic of $X_{2},X_{3},...,X_{M}$.
\item $G,$ is the distribution function of $Y_{1}$. $\forall y,\;G(y)=\left[F(y)\right]^{M-1}$.
\item $g=G',$ is the continuous density function of $G$ or $Y_{1}$.
\item $\Pi_{i},$ is the payoff of bidder $i$. $\Pi_{i}=\begin{cases}
x_{i}-b_{i}\quad if\;b_{i}>max_{j\neq i}b_{j}\\
0\qquad\quad if\;b_{i}<max_{j\neq i}b_{j}
\end{cases}$
\item $\Pi_{s},\;x_{s}$ is the payoff and valuation of the auction seller. 
\item $m\left(x\right),$ is the expected payment of a bidder with value
$x$.
\item $R_{s}$ is the expected revenue to the seller.
\item $\mathcal{M}=\left\{ 1,2,\ldots,M\right\} $ is the potential set
of bidders when there is uncertainty about how many interested bidders
there are.
\item $\mathcal{A}\subseteq\mathcal{N}$ is the set of actual bidders.
\item $p_{l}$ is probability that any participating bidder assigns to the
event that he is facing $l$ other bidders or that there is a total
of $l+1$ bidders, $l\in\left\{ 1,2,\ldots,M-1\right\} $. 
\item $X_{i}\in\left[0,\omega_{i}\right]$ is bidder $i's$ signal when
the valuations are interdependent. 
\item $V_{i}=\upsilon_{i}\left(X_{1},X_{2},...,X_{M}\right)$ is the value
of the exclusive to bidder $i$. $\upsilon_{i}\left(0,0,...,0\right)=0$
\item $\upsilon_{i}\left(x_{1},x_{2},...,x_{M}\right)\equiv E\left[V_{i}\mid X_{1}=x_{1},X_{2}=x_{2},...,X_{M}=x_{M},\right]$
is a more general setting, where knowing the signals of all bidders
still does not reveal the full value with certainty.
\end{itemize}

\section{\label{sec:Auction-Theory-Results}Appendix: Additional Auction Theory
Results}

The proofs for the below extensions are given in (Kashyap 2016).

\subsection{Asymmetric Valuations}

$f_{i},\,F_{i}$ , are the continuous density function and distribution
of bidder $i$ in this case where the valuations are asymmetric. $\phi_{i}\equiv\beta_{i}^{-1}$
is the inverse of the bidding strategy $\beta_{i}$. This means, $x_{i}=\beta_{i}^{-1}\left(b_{i}\right)=\phi_{i}\left(b_{i}\right)$.
The following result captures the scenario when we have an asymmetric
equilibrium.
\begin{lem}
\label{12-The-system-of}The system of differential equations for
an asymmetric equilibrium is given by
\begin{eqnarray*}
\sum_{j\neq i}^{j\in\left\{ 1,...,M\right\} }\left\{ \frac{f_{j}\left(\phi_{j}\left(b\right)\right)\phi'_{j}\left(b\right)}{F_{j}\left(\phi_{j}\left(b\right)\right)}\right\}  & = & \frac{1}{\left[\phi_{i}\left(b\right)-b\right]}
\end{eqnarray*}
\end{lem}
This system of differential equations can be solved to get the bid
functions for each player. Closed form solutions are known for the
case of uniform distributions with different supports. A simplification
is possible by assuming that say, some bidders have one distribution
and some others have another distribution. This is a reasonable assumption
since firms with bigger sources of internal inventory would tend to
differ from those with smaller sources. Among other things, this would
depend on the other divisions within a particular intermediary and
the reputation of its franchise. 
\begin{lem}
\label{13-If,--firms}If, $K+1$ firms (including the one for which
we derive the payoff condition) have the distribution $F_{1}$, strategy
$\beta_{1}$ and inverse function $\phi_{1}$. The other $M-K-1$
firms have the distribution $F_{2}$, strategy $\beta_{2}$ and inverse
function $\phi_{2}$. The system of differential equations is given
by,

\begin{eqnarray*}
\left\{ K\frac{f_{1}\left(\phi_{1}\left(b\right)\right)\phi'_{1}\left(b\right)}{\left[F_{1}\left(\phi_{1}\left(b\right)\right)\right]}\right\} +\left\{ \left(M-1-K\right)\frac{f_{2}\left(\phi_{2}\left(b\right)\right)\phi'_{2}\left(b\right)}{\left[F_{2}\left(\phi_{2}\left(b\right)\right)\right]}\right\} =\frac{1}{\left[\phi_{i}\left(b\right)-b\right]}
\end{eqnarray*}
\end{lem}
As a special case, if there are only two bidders, $M=2,K=1$ the above
reduces to a system of two differential equations, 
\begin{eqnarray*}
\phi'_{1}\left(b\right) & = & \frac{\left[F_{1}\left(\phi_{1}\left(b\right)\right)\right]}{f_{1}\left(\phi_{1}\left(b\right)\right)\left[\phi_{2}\left(b\right)-b\right]}\\
\phi'_{2}\left(b\right) & = & \frac{\left[F_{2}\left(\phi_{2}\left(b\right)\right)\right]}{f_{2}\left(\phi_{2}\left(b\right)\right)\left[\phi_{1}\left(b\right)-b\right]}
\end{eqnarray*}

\subsection{Symmetric Interdependent Valuations}

It is worth noting that a pure common value model of the sort, $V=\upsilon\left(X_{1},X_{2},...,X_{M}\right)$
is not entirely relevant in our context since the amount of internal
inventory and the size of the borrow book will vary across intermediaries.
This means that the amount of shares they will use from the exclusive
will vary and so will their valuations. What this reasoning tells
us is that it is reasonable to expect that there is some correlation
between the signals of each bidder. This makes sense since the total
supply, in a security, is distributed across all the bidders and the
valuation of the portion in the exclusive will depend on the total
supply. The valuation of a particular bidder will then depend on his
inventory and how he expects the rest of the supply to be distributed
among the other bidders, resulting in a symmetric interdependent auction
strategy. From the perspective of a particular bidder, the signals
of the other bidders can be interchanged without affecting the value.
This is captured using the function $u\left(X_{i},X_{-i}\right)$
which is the same for all bidders and is symmetric in the last $M-1$
components. We assume that all signals $X_{i}$ are from the same
distribution $\left[0,\omega\right]$ and that the valuations can
be written as 
\[
\upsilon_{i}\left(X_{1},X_{2},...,X_{M}\right)=u\left(X_{i},X_{-i}\right)
\]
We also assume that the joint density function of the signals $f$
defined on $\left[0,\omega\right]^{M}$ is symmetric and the signals
are affiliated. Affiliation here refers to the below properties.
\begin{itemize}
\item The random variables $X_{1},X_{2},...,X_{M}$ distributed on some
product of intervals $\mathcal{X}\subset\Re^{M}$ according to the
joint density function $f$. The variables $\mathbf{X=}\left(X_{1},X_{2},...,X_{M}\right)$
are affiliated if $\forall\mathbf{x',x''}\in\mathcal{X}$, $f\left(\mathbf{x'\lor x''}\right)f\left(\mathbf{x'\land x''}\right)\geq f\left(\mathbf{x'}\right)f\left(\mathbf{x''}\right)$.
Here $\mathbf{x'\lor x''}$ and $\mathbf{x'\lor x''}$ denote the
component wise maximum and minimum of $\mathbf{x'}$ and $\mathbf{x''}$.
\item The random variables $Y_{1},Y_{2},...,Y_{M-1}$ denote the largest,
second largest, ... , smallest from among $X_{2},X_{3},...,X_{M}$.
If $X_{1},X_{2},...,X_{M}$ are affiliated, then $X_{1},Y_{1},Y_{2},...,Y_{M-1}$
are also affiliated. 
\item Let $G\left(.\mid x\right)$ denote the distribution of $Y_{1}$ conditional
on $X_{1}=x$ and let $g\left(.\mid x\right)$ be the associated conditional
density function. Then if $Y_{1}$ and $X_{1}$ are affiliated and
if $x'>x$ then $G\left(.\mid x'\right)$ dominates $G\left(.\mid x\right)$
in terms of the reverse hazard rate, $\frac{g\left(t\right)}{G\left(t\right)}$.
That is $\forall y$,
\[
\frac{g\left(y\mid x'\right)}{G\left(y\mid x'\right)}\geq\frac{g\left(y\mid x\right)}{G\left(y\mid x\right)}
\]
\item If $\gamma$ is any increasing function, then $x'>x$ implies that
\[
E\left[\gamma\left(Y_{1}\right)\mid X=x'\right]\geq E\left[\gamma\left(Y_{1}\right)\mid X=x\right]
\]
\end{itemize}
We define the below function as the expectation of the value to bidder
1 when the signal he receives is $x$ and the highest signal among
the other bidders, $Y_{1}=y$. Because of symmetry this function is
the same for all bidders and we assume it is strictly increasing in
$x$. We also have $u\left(\mathbf{0}\right)=\upsilon(0,0)=0$. 
\[
\upsilon\left(x,y\right)=E\left[V_{1}\mid X=x,Y_{1}=y\right]
\]

\begin{lem}
\label{14-A-symmetric-equilibrium}A symmetric equilibrium strategy
governed by the set of conditions above is given by
\[
\beta\left(x\right)=\int_{0}^{x}\upsilon\left(y,y\right)dL\left(y\mid x\right)
\]
\end{lem}
Here, we define $L\left(y\mid x\right)$ as a function with support
$\left[0,\omega\right]$,
\[
L\left(y\mid x\right)=\exp\left[-\int_{y}^{x}\frac{g\left(t\mid t\right)}{G\left(t\mid t\right)}dt\right]
\]

\begin{lem}
\label{15-The-bidder's-equilibrium}The bidder's equilibrium strategy
under a scenario when the valuation is the weighted average of his
valuation and the highest of the other valuations is given by the
expression below. That is, we let $\upsilon\left(x,y\right)=\alpha x+\xi y\;\mid\alpha,\xi\in\left[0,1\right]$.
This also implies, $\upsilon\left(x,y\right)=u\left(x,y\right)=u\left(x_{i},x_{-i}\right)=\alpha x_{i}+\xi\max\left(x_{-i}\right)$,
giving us symmetry across the signals of other bidders. An alternative
formulation could simply be $\upsilon\left(x,y\right)=\frac{1}{M}\left(\overset{M}{\underset{i=1}{\sum}}x_{i}\right)$.
The affiliation structure follows the Irwin-Hall distribution (End-note
\ref{Irwin-Hall}) with bidder's valuation being the sum of a signal
coming from a uniform distribution with $\omega=1$ and a common component
from the same uniform distribution. 
\end{lem}
\begin{eqnarray*}
\beta\left(x\right) & = & \left[\frac{2\left(\alpha+\xi\right)\left(M-1\right)}{\left(2M-1\right)x^{2M-2}}\right]\\
 & + & \left(\alpha+\xi\right)\left[x-\frac{1}{\left(2x-1-\frac{x^{2}}{2}\right)^{M-1}}\left\{ \frac{1}{2^{M-1}}+\int_{1}^{x}\left(2y-1-\frac{y^{2}}{2}\right)^{M-1}dy\right\} \right]
\end{eqnarray*}

\subsection{Combined Realistic Setting}
\begin{lem}
\label{16-The-bidding-strategy}The bidding strategy in a realistic
setting with symmetric interdependent, uniformly distributed valuations,
with reserve prices and variable number of bidders is given by 
\end{lem}
\[
\beta\left(x\right)=re^{-\int_{x^{*}}^{x}\frac{g\left(t\mid t\right)}{G\left(t\mid t\right)}dt}+\int_{x^{*}}^{x}v(y,y)\frac{g\left(y\mid y\right)}{G\left(y\mid y\right)}e^{-\int_{y}^{x}\frac{g\left(t\mid t\right)}{G\left(t\mid t\right)}dt}dy
\]

Here, $x^{*}\left(r\right)$ is found by solving for $x$ in the below
condition 
\[
\left[\int_{0}^{1}\xi y\left[\frac{y}{x}\right]^{2\left(M-2\right)}\left(\frac{2y}{x^{2}}\right)dy+\int_{1}^{x}\xi y\left[\frac{\left(2y-1-\frac{y^{2}}{2}\right)}{\left(2x-1-\frac{x^{2}}{2}\right)}\right]^{M-2}\left\{ \frac{2-y}{\left(2x-1-\frac{x^{2}}{2}\right)}\right\} dy\right]=\frac{r-\alpha x}{\left(M-1\right)}
\]

It is trivial to extend the above to the case where the total number
of bidders is uncertain by using the equilibrium bidding strategy
$\beta^{l}\left(x\right)$ and the associated probability $p_{l}$
when there are exactly $l+1$ bidders, known with certainty, 
\[
\beta^{M}\left(x\right)=\sum_{l=0}^{M-1}\frac{p_{l}G^{l}\left(x\right)}{G\left(x\right)}\beta^{l}\left(x\right)
\]

\end{document}